\documentclass[pagesize]{scrartcl}
\usepackage[pdftex]{graphicx}
\usepackage{epstopdf}

\usepackage[mathlines]{lineno}

\usepackage{setspace}

\newcommand*\patchAmsMathEnvironmentForLineno[1]{%
  \expandafter\let\csname old#1\expandafter\endcsname\csname #1\endcsname
  \expandafter\let\csname oldend#1\expandafter\endcsname\csname end#1\endcsname
  \renewenvironment{#1}%
     {\linenomath\csname old#1\endcsname}%
     {\csname oldend#1\endcsname\endlinenomath}}%
\newcommand*\patchBothAmsMathEnvironmentsForLineno[1]{%
  \patchAmsMathEnvironmentForLineno{#1}%
  \patchAmsMathEnvironmentForLineno{#1*}}%
\AtBeginDocument{%
\patchBothAmsMathEnvironmentsForLineno{equation}%
\patchBothAmsMathEnvironmentsForLineno{align}%
\patchBothAmsMathEnvironmentsForLineno{flalign}%
\patchBothAmsMathEnvironmentsForLineno{alignat}%
\patchBothAmsMathEnvironmentsForLineno{gather}%
\patchBothAmsMathEnvironmentsForLineno{multline}%
}

\usepackage[scaled=.90]{helvet}
\usepackage{mathptmx}
\usepackage{courier}

\usepackage{natbib}
\usepackage[ngerman,english]{babel}
\usepackage{dsfont}

\usepackage{amsmath}
\numberwithin{equation}{section}

\usepackage{amssymb}

\usepackage{enumerate, amssymb, amsmath, amsthm}

\usepackage{tikz} 
\usepackage[T1]{fontenc}
\usepackage{lmodern}
\usepackage[labelfont={bf,it},labelsep=period]{caption}
\usepackage{units}

\usepackage{changes}

\definechangesauthor[name={Soeren}, color=red]{soe}
\definechangesauthor[name={Sebastian}, color=blue]{seb}

\usepackage{ifthen}

\newcommand*{\appendixmore}{%
  \renewcommand*{\othersectionlevelsformat}[1]{%
    \ifthenelse{\equal{##1}{section}}{\appendixname~}{}%
    \csname the##1\endcsname\autodot\enskip}
  \renewcommand*{\sectionmarkformat}{%
    \appendixname~\thesection\autodot\enskip}
}

\newtheorem{satz}{Theorem}[section]

\newtheorem{koro}[satz]{Corollary}

\newtheorem{bsp}[satz]{Example}

\newcommand{\R}{\ensuremath{{\mathbb R}}}
\newcommand{\E}{\ensuremath{{\mathbb E}}}

\DeclareMathOperator*{\sgn}{sgn}

\begin{document}

\title{Anscombe's Model for Sequential Clinical Trials Revisited\footnote{\textbf{Acknowledgement.} This project has received funding from the European Union's 7th Framework Programme for research, technological development and demonstration under the IDEAL Grant Agreement no 602552. We are very grateful to Carl-Fredrik Burman, Christopher Jennison and Stephen Senn for valuable discussions during the research project.}
}

\author{  
  Sebastian Jobj\"{o}rnsson\thanks{
    Department of Mathematical Sciences, Chalmers University of Technology, SE-412 96, Gothenburg, Sweden. E-mail: jobjorns@chalmers.se.},
  S\"{o}ren Christensen\thanks{
    \textbf{Corresponding author.} Department of Mathematics, Research Group Statistics and Stochastic Processes, University of Hamburg, Bundesstr.\ 55 (Geomatikum), 20146 Hamburg, Germany. E-mail: soeren.christensen@uni-hamburg.de}
}

\date{\today}
\maketitle
\begin{center}\textbf{Short title (for the running head): Anscombe's Model Revisited} \end{center}

\begin{abstract}
  In Anscombe's classical model, the objective is to find the optimal sequential rule for learning about the difference between two alternative treatments and subsequently selecting the superior one. The population for which the procedure is optimised has size $N$ and includes both the patients in the trial and those which are treated with the chosen alternative after the trial. We review earlier work on this problem and give a detailed treatment of the problem itself. In particular, while previous work has mainly focused on the case of conjugate normal priors for the incremental effect, we demonstrate how to handle the problem for priors of a general form. We also discuss methods for numerical solutions and the practical implications of the results for the regulation of clinical trials.

  Two extensions of the model are proposed and analysed. The first breaks the symmetry of the treatments, giving one the role of the current standard being administered in parallel with the trial. We show how certain asymptotic results due to Chernoff can be adapted to this asymmetric case. The other extension assumes that $N$ is a random variable instead of a known constant.
  \vspace{.4cm}
  ~
\end{abstract}

\thispagestyle{empty}

\clearpage
\setcounter{page}{1}

\section{Introduction}
In an extended review \citep{Anscombe1963} of Peter Armitage's book \emph{Sequential Medical Trials} \citep{Armitage1960}, Anscombe discusses a classical decision-theoretic model for sequential clinical trials. In this model two treatments, $A$ and $B$, are compared in a sequential trial. The goal is to find an optimal rule for determining when to stop the trial and which of the two treatments that should then be given to the remaining patient population. In order to make the discussion of the central issues as clear as possible, Anscombe makes a number of simplifying assumptions. We shall follow him in this. In particular, we assume that the two treatments are indistinguishable in terms of costs and side effects. Hence, the data provided by the trial is used solely for decreasing the uncertainty regarding the incremental effect of $A$ relative to $B$.

The present paper gives details on the underlying mathematics for solving the original problem and proposes two different types of model extensions. The first extension follows from the observation that, in practice, it is unrealistic that all patients in the target population participate in the trial phase. Typically, those that are not part of the trial will still receive some treatment (the current standard). We demonstrate that, if it assumed that this treatment is $B$, then a new optimal stopping problem can be formulated and solved with the same method used for Anscombe's original problem. In the other extension it is assumed that the total number of patients is no longer fixed, but instead considered to be a random variable. The problem is identified to be of a similar form as before. For some particular cases, the structure of the problem simplifies substantially, allowing for explicit solutions. To the best of the authors' knowledge, these are the first explicit solutions in this context.

We now proceed to the formulation of Anscombe's original problem. Assume that each of $N$ patients will be treated by either treatment $A$ or $B$. This total number of patients is referred to as the patient horizon and includes the subjects enrolled in the trial. Given that $X_A$ and $X_B$ denote the responses for a patient given $A$ and $B$, respectively, we assume conditional normal distributions according to
\begin{equation*}
  X_A \sim \mathcal{N} \left( \theta + \mu, \sigma^2 \right), \quad X_B \sim \mathcal{N} \left( \theta - \mu, \sigma^2 \right).
\end{equation*}
Here, $\theta$ denotes the mean effect of $A$ and $B$ in the population, while $2 \mu$ is the incremental effect of $A$ relative to $B$. Hence, $\mu > 0$ means that treatment $A$ is preferred to treatment $B$ and vice versa. The common variance parameter $\sigma^2$ is assumed to be known. For a sample size of $n$ patients, it is assumed that half are given $A$ and half are given $B$. Let $\Sigma^A_n$ and $\Sigma^B_n$ denote the response sums for patients given $A$ and $B$, respectively. Under the assumption of i.i.d.\ normal responses given $\theta$ and $\mu$, we have
\begin{equation*}
  \Sigma^A_n \sim \mathcal{N} \left( \frac{n}{2}(\theta + \mu), \frac{n}{2} \sigma^2 \right), \quad \Sigma^B_n \sim \mathcal{N} \left( \frac{n}{2} (\theta - \mu), \frac{n}{2} \sigma^2 \right) .
\end{equation*}
Letting the difference in response sums be denoted by $\Sigma_n$, it then follows that, given $\mu$,
\begin{equation*}
  \Sigma_n = \Sigma^A_n - \Sigma^B_n \sim \mathcal{N} \left( n \mu, n \sigma^2 \right) .
\end{equation*}

Based on the initial beliefs about the unknown effect parameter $\mu$, the trial sample size $n$ and the observed response difference $\Sigma_n$, either $A$ or $B$ will be chosen for all the $N - n$ remaining patients. In practice, confirmatory (phase III) trials often have a pre-determined sample size or use a group-sequential design with very few (one or two, say) interim analyses. However, we will assume that the experiment is fully sequential. Hence, the responses of each additional patient pair can be observed immediately and we may stop the trial after any (even) number of observations.

The goal is to find an optimal rule $(\tau, D_\tau)$. The first part, $\tau$, denotes a stopping rule, mapping each pair $(n, \Sigma_n)$ into a decision of whether to stop the trial or continue taking samples. The second part denotes the treatment decision given that the trial is stopped, with $D_\tau = 1$ corresponding to treatment $A$ and $D_\tau = -1$ corresponding to treatment $B$. Using $\tau$ as a stopping rule, the total number of superior and inferior treatments is $\tau/2$ each in the trial phase, so that the expected response is $\theta \tau$ for the trial patients. For the post-trial patients, the expected response is $(N - \tau)(\theta + D_\tau \mu)$. Hence, the total expected response for a given $\mu$ over the patient horizon is $N \theta + \mu \E^\mu \left[ (N - \tau) D_\tau \right]$, where $\E^\mu$ denotes the expectation given a specific value of $\mu$. Note that the leading term $N \theta$ can be ignored when optimising, since it does not depend on the decision rule. Therefore, for a given value of $\mu$ and a rule $(\tau, D_\tau)$, we are faced with the expected utility
\begin{equation} \label{eq:utility}
  \mu \E^\mu \left[(N - \tau) D_\tau \right] .
\end{equation}
In principle, having specified a prior distribution for $\mu$, it is possible to solve the resulting optimal stopping problem by straightforward backward induction. However, even for moderate values of $N$ the required computations will be quite heavy. Noting that $\Sigma_n$ is a Gaussian random walk with $\E \left[ \Sigma_n \right] = n \mu$ and $\text{Var}(\Sigma_n) = n \sigma^2$, we can naturally extend the discrete time process and instead consider a Brownian motion $(\Sigma_t)_{t \in [0, N]}$ with drift $\mu$, volatility $\sigma$ and filtration $(\mathcal{F}_t)_{t\in[0,N]}$. Maximisation of the expected value of \eqref{eq:utility}, with respect to a prior for $\mu$, then leads to a Markovian optimal stopping problem that can be solved using the free-boundary approach.

An overview of related literature is given in the following subsection. We begin our analysis by presenting a framework for solving the problem for a general prior distribution in Section \ref{sec:bayes}. Using Girsanov's transform, it is shown that the problem is equivalent to an ordinary Markovian optimal stopping problem. A standardising transformation then leads to a problem which is independent of the parameters $\sigma$ and $N$. A non-linear integral equation defining the unique optimal stopping boundary in the special case of a symmetric prior distribution is derived. The special case of conjugate normal priors is studied in Section \ref{sec:normal_free_bound}. It turns out that one really only needs to solve a single optimal stopping problem in order to obtain the optimal stopping boundary for any prior in this class. Section \ref{sec:maximin} takes a maximin approach to the problem and shows that this is equivalent to placing a symmetric two-point prior on the unknown effect parameter. Numerical schemes available for solving the integral equations are discussed in Section \ref{sec:solving_numerically}. The first of our extensions to Anscombe's problem is treated in Section \ref{sec:asymmetric_problem}. It is shown that this model leads to asymmetric stopping boundaries which, as for the original problem, can be described using integral equations. Certain asymptotic results first derived by \citet{Chernoff1981}, adapted to this asymmetric case, are presented in Section \ref{subsec:asymptotic_results}. The other extension, with a random patient horizon, is presented in Section \ref{sec:random_nb}. A discussion of the practical applicability of Anscombe's model is given in Section \ref{sec:discussion}, where we also consider its implications for the regulation of clinical trials. Limitations and possible further extensions of the model are also considered.

\subsection{Related literature} \label{subsec:related_literature}
One of the first major results in what would become the subject area of sequential analysis was obtained by Wald \citep{Wald1945,WW1}. Essentially, the classical Neyman-Pearson approach for constructing a statistical test of a null hypothesis $H_0$ against a single alternative $H_1$ consists of selecting the non-sequential test with the minimum sample size required to achieve a certain type II error from a class of tests in which all members have the same type I error. Wald showed that a reduction in the expected sample size could be achieved by allowing for sequential analysis of the data. In particular, he demonstrated the construction of a decision rule called the sequential probability ratio test (SPRT), which minimises the expected sample size among all sequential rules satisfying fixed requirements on the type I and type II errors.

Wald's SPRT is for all practical purposes the most efficient sequential test when both $H_0$ and $H_1$ are simple hypotheses, that is, when they both consist of single points in the parameter space. However, the fully Bayesian decision-theoretic perspective calls for a more flexible methodology. Instead of single point hypotheses, one would like to be able to specify a prior distribution on the whole of the parameter space. Further, minimisation of expected sample size under type I and type II error restrictions may not be the appropriate objective when selecting the decision rule. Instead, the decision-maker would typically want to find the optimal rule with respect to a utility function that captures all the gains and costs involved with each possible sequence of decisions.

A large part of the early methodology for solving optimal stopping problems using the free-boundary approach was developed by Chernoff in a series of publications during the 1960s \citep{Chernoff1961,BreakwellChernoff1964,Chernoff1965}. The problem studied in these papers is different from the one treated here. It involves the sequential testing of whether the mean of a normal distribution is positive or negative, given an infinite time horizon and a fixed cost per unit of observation time. Various properties of the optimal boundary for the problem studied in these papers are given. In particular, asymptotic expansions as the observation time goes to 0 and infinity are derived. However, the optimal boundary is not characterised over the entire time horizon. In a more recent publication by \citet{ZhitlukhinMuravlev2013}, an integral equation is derived for a transformed version of the optimal boundary. This equation can then be solved numerically in order to obtain a full description.   

The particular optimal stopping problem investigated in this paper was formulated by \citet{Anscombe1963}. He aims to minimise expected regret rather than to maximise an explicit expression for the total expected utility, but since he assumes that the regret is proportional to the absolute value of the unknown, incremental effect size, the resulting mathematical model is the same as ours. In \citep{Chernoff1981}, the free-boundary methodology developed earlier was applied to the problem proposed by Anscombe. After making a standardising transformation, they derive (numerically) the optimal procedure in the case of a normal conjugate prior on the unknown incremental efficacy and compare its performance with the approximate solution suggested by Anscombe and another alternative suggested by \citet{BeggMehta1979}. They also characterise the asymptotic behaviour of the optimal stopping boundary for small sample sizes, and provide a proof in the appendix of the paper. This proof has been of particular value to us when deriving corresponding asymptotics for the asymmetric generalisation of Anscombe's problem in which treatment allocation may be unbalanced during the trial phase.

A common criticism often raised in connection with Anscombe's model is that it does not reflect ethical considerations enough. This was a motivation for extending the model by introducing ethical costs \citep{Chernoff1985}. \citet{Petkau1987} solved the continuous time problem with a normal prior under the constraint that the trial must stop before reaching a fixed fraction of the horizon $N$. In \citep{Petkau2003}, optimal group sequential designs for Anscombe's model are derived for both the truncated and original version of the problem.

A recent contribution by \citet{Stallard2017} solves the problem of sample size optimisation when comparing two treatments in a non-sequential trial. In contrast to the normal response and specific gain function assumed in Anscombe's model, their framework allows for response distributions of general exponential family form depending on a single parameter. Further, arbitrary continuous gain functions of the parameter may be specified. The main result obtained is that the optimal fixed sample sizes are $O(\sqrt{N})$ as $N \to \infty$.

Although this literature review is by necessity quite selective, it shows that there has been a continuous interest in Anscombe's problem in the statistical community throughout the decades since it was first introduced. However, the modern theory and results surrounding the free-boundary approach to solving optimal stopping problems that we employ here is not tied to this specific setting but has a much wider applicability. A comprehensive reference work is the monograph by \citet{ps}, which applies the theory to a range of different problems in stochastic analysis, statistics and finance. For a survey covering the origins of the modern approach and more recent applications of the theory to problems in different areas, see the article by \citet{Lai2005}. In addition to discussing applications to problems of sequential analysis and Bayes optimisation of clinical trials, Lai and Lim also review the work done by Bather and Chernoff on singular stochastic control in the 1960's, and their own contributions to problems in option pricing hedging in the presence of transaction costs.

\section{Bayesian formulation with general prior distributions} \label{sec:bayes}
Given a prior distribution $\nu$ for $\mu$ in formula \eqref{eq:utility}, the problem to solve is
\begin{equation} \label{eq:general_bayes_problem}
  \sup_{\substack{0 \le \tau \le N \\ D_\tau \in \{ 1, -1\}}} \int \! \mu \E^\mu \left[ (N - \tau) D_\tau \right] \, \nu( \mathrm{d} \mu) .
\end{equation}
We tackle this problem by first applying Girsanov's transform \citep[Theorem 8.6.4]{Oks}. Note that this is not a new approach in these kinds of problems, but corresponds to standard methodology in filtering theory, see \citet[Section 3.3]{bain2009fundamentals}. For example, \citet{Ekstrom2015} make use of essentially the same technique when analysing a sequential testing problem of the drift of a Brownian motion. If a change of measure is defined via
\begin{equation}
  \frac{\mathrm{d} \mathbb{P}|_{\mathcal{F}_t}}{\mathrm{d} \mathbb{P}^{\mu}|_{\mathcal{F}_t}}=\exp\left(-\frac{\mu}{\sigma^2} \Sigma_t + \frac{\mu^2}{2 \sigma^2} t \right),
\end{equation}
then the process $(\Sigma_t)$ is a Brownian motion with drift 0, volatility $\sigma$ and starting value $\Sigma_0 = 0$ under the new measure $\mathbb{P}$. Making use of this in problem \eqref{eq:general_bayes_problem} we obtain that for each decision rule $(\tau, D_\tau)$,
\begin{align*}
  & \int \! \mu \E^\mu \left[ (N - \tau) D_\tau \right] \, \nu( \mathrm{d} \mu) = \int \! \mu \E \left[ \exp \left( \frac{\mu}{\sigma^2} \Sigma_\tau - \frac{\mu^2}{2\sigma^2} \tau \right) (N - \tau) D_\tau \right] \, \nu( \mathrm{d} \mu) = \\
& \E \left[ \int \! \mu \exp \left(\frac{\mu}{\sigma^2} \Sigma_\tau - \frac{\mu^2}{2\sigma^2} \tau \right) (N - \tau) D_\tau \, \nu(\mathrm{d} \mu) \right] = \E \left[ h_\nu(\tau / \sigma^2, \Sigma_\tau / \sigma^2) (N - \tau) D_\tau \right], 
\end{align*}
\begin{equation} \label{eq:h_function_def}
  \text{where} \quad h_\nu (t,x) \equiv \int \! \mu \exp \left( \mu x - \frac{\mu^2}{2} t \right) \, \nu(\mathrm{d} \mu).
\end{equation}
For the problem to be nontrivial, we assume here and in the following that $\nu$ has finite first moment, which guarantees that the function $h_\nu$ is well-defined for $(t,x) \in (0, N] \times \R$.

  Let $\sgn(\cdot)$ denote the function
  \begin{equation*}
    \sgn (x) = \begin{cases}
      1, & \quad x \ge 0, \\
      -1, & \quad x < 0 .     
    \end{cases}
  \end{equation*}
  For a fixed stopping time $\tau$, it is then clear from the calculation above that the expectation in problem \eqref{eq:general_bayes_problem} is maximised for $D^*_\tau = \sgn \left( h_\nu(\tau / \sigma^2, \Sigma_\tau / \sigma^2) \right)$. To get the optimal expected utility, it therefore remains to maximise the expectation
\begin{equation*}
  \E \left[ (N - \tau) \big| h_\nu(\tau / \sigma^2, \Sigma_\tau / \sigma^2) \big| \right] .
\end{equation*}
Note that this is a standard Markovian stopping problem. We have now proved
\begin{satz}\label{thm:bayes_osp}
In the optimisation problem \eqref{eq:general_bayes_problem} with a prior $\nu$, the optimal decision variable is given by $D^*_\tau = \sgn \left( h_\nu(\tau / \sigma^2, \Sigma_\tau / \sigma^2) \right)$ and the optimal stopping time $\tau^*$ solves
\begin{equation} \label{eq:OSP}
  \sup_{0\leq \tau\leq N} \E \left[ (N - \tau) \big| h_\nu(\tau / \sigma^2, \Sigma_\tau / \sigma^2) \big| \right].
\end{equation}
\end{satz}

In order to conveniently exploit the Markovian structure in optimal stopping problems of type \eqref{eq:OSP}, the canonical first step is to extend the problem to general starting states $(t, x)$. In the present case, we do this by introducing a value function $V(t,x)$ and a reward function $G(t,x)$ according to 
\begin{align}
  V(t, x) & = \sup_{t \le \tau \le N} \mathbb{E}_{(t, x)} \left[ G(\tau, \Sigma_\tau) \right], \label{eq:cont_stopping_problem} \\
  G(t, x) & = (N - t) \big| h_\nu(t / \sigma^2, x / \sigma^2) \big| . \label{eq:cont_reward} 
\end{align}
Here and in the following, $\mathbb{E}_{(t, x)}$ denotes the expectation given that the process is started in point $(t, x)$. The original problem, as formulated in \eqref{eq:OSP}, then corresponds to the starting state $(t,x)=(0,0)$.

\subsection{A standardising transformation} \label{subsec:standardising_transformation}
It is convenient to make a time-space transformation in the problem defined by equations \eqref{eq:cont_stopping_problem} and \eqref{eq:cont_reward} before attempting to solve it. Define $\delta \equiv \mu \frac{\sqrt{N}}{\sigma}$. The prior distribution $\nu$ for $\mu$ then corresponds to a prior distribution $\xi$ for $\delta$ satisfying $\xi(E) = \nu \left(\frac{\sigma}{\sqrt N} E \right)$ for any measurable subset $E$ of the real line. Further, define 
\begin{equation*}
  r = \frac{t}{N}, \quad y = \frac{x}{\sqrt{N \sigma^2}}, \quad S_r = \frac{\Sigma_{Nr}}{\sqrt{N \sigma^2}} .
\end{equation*}
By Brownian scaling, this makes the new process $(S_r)$ a standard Brownian motion. The function $h_\nu$ can now be rewritten as
\begin{align*}
  h_\nu (\tau / \sigma^2, \Sigma_\tau / \sigma^2) & = \int \! \mu \exp \left( \frac{\mu}{\sigma^2} \Sigma_\tau - \frac{\mu^2}{2 \sigma^2} \tau \right) \, \nu(\mathrm{d} \mu) = \int \! \mu \exp \left( \frac{\mu}{\sigma}\sqrt{N} S_{\tau/N} - \frac{\mu^2}{2\sigma^2} \tau \right) \, \nu(\mathrm{d} \mu) \\
  & = \frac{\sigma}{\sqrt{N}} \int \! \delta \exp \left( \delta S_{\tau/N} - \frac{\delta^2}{2} (\tau / N) \right) \, \xi(\mathrm{d} \delta)  = \frac{\sigma}{\sqrt{N}} h_\xi \left(\tau / N, S_{\tau/N} \right) .
\end{align*}
It follows that 
\begin{equation*}
  G(\tau, \Sigma_\tau) = \sigma \sqrt{N} (1 - \tau / N) \left| h_\xi \left(\tau / N, S_{\tau/N} \right) \right| .
\end{equation*}
Since the constant factor $\sigma \sqrt{N}$ does not affect the optimal solution, the original problem is thus reduced to the standardised form
\begin{align}
  \tilde{V}(r, y) & = \sup_{r \le \rho \le 1} \mathbb{E}_{(r, y)} \left[ \tilde{G}(\rho, S_\rho) \right], \label{eq:cont_stopping_problem_transformed} \\
  \tilde{G}(r, y) & = (1 - r) |h_\xi (r, y)| \label{eq:cont_reward_transformed} .
\end{align}

\subsection{Symmetric priors} \label{subsec:symmetric_prior}
Having shown how Anscombe's problem can be formulated as a standard Markovian optimal stopping problem for general priors, we now turn to the question of how to obtain solutions using the method of integral equations. Assuming sufficient regularity, these solutions can be expressed as boundaries enclosing a continuation region for the observed process. In general, the continuation region for the process need not be symmetric if priors of a general form are allowed. In order to simplify the presentation of the central results, we restrict the theoretical development in this section to the case of symmetric priors, satisfying $\xi(\mathrm{d} \delta) = \xi(-\mathrm{d} \delta)$. In this case, it is immediately seen that
\begin{equation} \label{eq:symmetric_h_function}
  h_\xi (r, y) = 2 \int_0^\infty \! \delta \exp \left(-\frac{\delta^2}{2} r \right) \sinh \left( \delta y \right) \, \xi(\mathrm{d} \delta) .
\end{equation}
Thus, $h_\xi(r, \cdot)$ becomes an odd function for fixed $r$. We therefore obtain
\begin{koro} \label{cor:symm}
In the optimisation problem defined by equations \eqref{eq:cont_stopping_problem_transformed} and \eqref{eq:cont_reward_transformed} with a symmetric prior $\xi$, the optimal decision variable is given by $D^*_\rho = \sgn (S_\rho)$ and the optimal stopping time $\rho^*$ solves the Markovian optimal stopping problem
\begin{align*}
  \sup_{0 \leq \rho \leq 1} \E \Big[ (1 - \rho) h_\xi \left( \rho, \left| S_\rho \right| \right) \Big] .
\end{align*}
\end{koro}

By the general theory of optimal stopping problems for Markov processes, the optimal stopping time may be expressed as
\begin{equation*}
  \rho^* = \inf\{r : (r, S_r) \in \mathbb{S} \},~\text{where } \mathbb{S} \equiv \{(r,y) : \tilde V(r,y) = \tilde G(r,y) \} .
\end{equation*}
Since the prior $\xi$ is symmetric, it is clear that the optimal stopping set $\mathbb{S}$ must also be symmetric. Now, our interest here is not in providing the precise technical conditions under which a solution may be obtained via the free-boundary method and integral equations. Detailed discussions are provided for a long list of similar problems elsewhere in the literature, see \cite{ps} for an overview or \citet{Ekstrom2015} for a more recent result. Rather, we wish to illustrate how the general approach may be used to solve Anscombe's classical problem and various extensions of it. Therefore, it will be assumed in what follows that the prior $\xi$ is sufficiently regular to imply the existence of a continuous boundary function $b_S(r)$ such that the stopping time $\rho^* = \inf \{r : |S_r| \ge b_S(r) \} \wedge 1$ is optimal. Following the arguments of \citet{Pedersen2002}, an application of a generalised version of It\^{o}'s formula then --- under weak assumptions --- leads to the following integral equation for $b_S$:
\begin{multline} \label{eq:symmetric_free_boundary_equation}
  \mathbb{E}_{(r, b_S(r))} \left[ \tilde{G} (1, S_1) \right] = \tilde{G} (r, b_S(r)) \\ + \int_r^1 \! \mathbb{E}_{(r, b_S(r))} \left[ \left( \frac{\partial}{\partial r} + \frac{1}{2} \frac{\partial^2}{\partial^2 y} \right) \tilde{G} (u, S_u) \, \mathbb{I} \left( | S_u | \ge b_S (u) \right) \right] \, \mathrm{d}u , \quad 0 \le r \le 1.
\end{multline}
Since $\tilde{G}$ vanishes when $r = 1$, the left hand side of equation \eqref{eq:symmetric_free_boundary_equation} is zero. The differential operator applied to $\tilde{G}$ gives
\begin{equation*}
  \left( \frac{\partial}{\partial r} + \frac{1}{2} \frac{\partial^2}{\partial^2 y} \right) \tilde{G} (r, y) = - h_\xi(r, |y|) .
\end{equation*}
Hence, the integral equation is reduced to
\begin{equation} \label{eq:symmetric_free_boundary_equation_reduced}
(1 - r) h_\xi(r, b_S(r)) = \int_r^1 \! \mathbb{E}_{(r, b_S(r))} \Big[ h_\xi (u, |S_u|) \, \mathbb{I} \left( | S_u | \ge b_S (u) \right) \Big] \, \mathrm{d}u, \quad 0 \le r \le 1.
\end{equation}

The expectation in equation \eqref{eq:symmetric_free_boundary_equation_reduced} can be rewritten in terms of the standard normal distribution function $\Phi$ as
\begin{equation} \label{eq:symmetric_expectation_normal_CDF}
  \int_{-\infty}^\infty \! \delta e^{-\delta^2 r / 2} \left\{ e^{\delta b_S(r)} \Phi \left( A_\delta^+ \right) + e^{- \delta b_S(r)} \Phi \left( A_\delta^- \right) \right\} \, \xi(\mathrm{d}\delta),
\end{equation}
where
\begin{equation*}
  A_\delta^\pm \equiv \frac{\delta (u - r) - b_S(u) \pm b_S(r)}{\sqrt{u - r}} .
\end{equation*}

A result by \citet[Theorem 2.1]{Pedersen2002}, slightly adapted to our setting, leads to the following uniqueness result for the solution to equation \eqref{eq:symmetric_free_boundary_equation_reduced}:
\begin{satz} \label{thm:integral_equation_uniqueness}
  Suppose $\xi$ is a symmetric measure and sufficiently regular to imply the existence of a continuous $b_S(r)$ satisyfing equation \eqref{eq:symmetric_free_boundary_equation_reduced} and making $\inf \{r : |S_r| \ge b_S(r) \} \wedge 1$ an optimal stopping time. Then $b_S(r)$ is the unique solution to equation \eqref{eq:symmetric_free_boundary_equation_reduced} in the class of all continuous boundaries.
\end{satz}
The proof of Theorem \ref{thm:integral_equation_uniqueness} is given in Appendix \ref{sec:integral_equation_uniqueness}.

\section{Conjugate normal priors} \label{sec:normal_free_bound}
We now illustrate how Theorem \ref{thm:bayes_osp} (after the standardising transformation) can be used to obtain the optimal boundary for a normal prior, which is conjugate to the Brownian motion model assumed for the sum process $(S_r)$. Given a prior for $\mu$ of the form $\nu = \mathcal{N}(v_0, \sigma_0^2)$, the corresponding prior for $\delta$ after the transformation $\delta = \mu \frac{\sqrt{N}}{\sigma}$ is
\begin{equation*}
  \xi = \mathcal{N} \left( m_0 \equiv v_0 \frac{\sqrt{N}}{\sigma}, \frac{N}{\sigma^2} \sigma_0^2 \right) .
\end{equation*}
A convenient parameterisation of the prior variance is now obtained by letting $n_0$ be the number of 'prior observations' and defining $\sigma_0^2 = \sigma^2 / n_0$. Letting $r_0$ be the ratio $n_0 / N$, it follows that $\xi = \mathcal{N}(m_0, 1 / r_0)$.

Up to a constant that doesn't play a role for the optimisation, elementary calculus yields that
\begin{equation*}
  h_\xi (r, y) = m_r(y) \beta(r, y), \quad m_r(y) \equiv \frac{m_0 r_0 + y}{r_0 + r}, \quad \beta(r, y) \equiv \frac{\exp \left( \frac{1}{2} m_r(y)^2 (r_0 + r) \right)}{\sqrt{r_0 + r}} .
\end{equation*}
Here, $m_r(y)$ is the mean of the posterior distribution for $\delta$ at time $r$ given that $S_r = y$. The important observation for treating normal priors with arbitrary parameters at the same time is that the function $\beta$ is a positive time-space harmonic function in the sense that the process $( \beta(r, S_r) )$ is a martingale. This implies that $\beta$ can be used to perform a change of measure using the $h$-transform \citep[II.31]{Borodin}. Under the new measure $\tilde{\mathbb{P}}$ with corresponding expectation $\tilde{\mathbb{E}}$ we have
\begin{equation*}
  \E \Big[ (1 - \rho) \big| h_\xi (\rho, S_\rho) \big| \Big] = \tilde \E \Big[ (1 - \rho) |M_\rho| \Big],
\end{equation*}
where $M_r \equiv m_r(S_r)$ is the posterior mean process. Using standard facts about the $h$-transform, it is immediately checked that $(M_r)$ is a diffusion process with zero drift starting in $(r, y) = (0, m_0)$. Whenever we work with this process, we assume that we work under the probability measure $\tilde{\mathbb{P}}$. As we do not expect it to cause any confusion, we leave out the tilde in the following in order to simplify the exposition.

Every prior that is conjugate normal is defined by the two parameters $m_0$ and $r_0$. At a first glance, it would therefore seem reasonable to expect that one would have to solve a distinct optimal stopping problem for each such pair $(m_0, r_0)$. However, as we will now demonstrate, by using a certain time transformation, the optimal stopping problem for $(M_r)$ may be transformed into a new one for which the optimal boundary is independent of the prior parameters. The solution for a particular pair $(m_0, r_0)$ can then be recovered by a simple transformation of the optimal boundary for the standardised problem. Specifically, introduce a new time $s$ via the transformation
\begin{equation} \label{eq:time_transformation}
s = - \frac{r_0 + 1}{r_0 + r} \iff r = - \left( r_0 + \frac{r_0 + 1}{s} \right) ,
\end{equation}
implying that $r \in [0, 1]$ corresponds to $s \in \left[ -1 - r_0^{-1}, -1 \right]$. Applying this transformation to the reward function obtained after the $h$-transform gives
\begin{equation*}
  (1 - r) |M_r| = \left( 1 + \left( r_0 + \frac{r_0 + 1}{s} \right) \right) |M_{r(s)}| = \sqrt{r_0 + 1} \left(1 + s^{-1} \right) \left| \sqrt{r_0 + 1} \, M_{r(s)} \right| .
\end{equation*}
The time change is chosen so that the process $W_s \equiv \sqrt{r_0 + 1} \, M_{r(s)}$ is a standard Brownian motion (see, e.g., \citet[Corollary 8.5.3]{Oks}). Since a constant factor of $\sqrt{r_0 + 1}$ does not affect the optimal stopping time, the standardised problem for the new time $s$ is
\begin{align}
\hat{V} (s, y) & = \sup_{s \le \zeta \le -1} \E_{(s, y)} \left[ \hat{G} (\zeta, W_\zeta) \right], \label{eq:cont_stopping_problem_trans} \\
\hat{G} (s, y) & = \left(1 + s^{-1} \right) |y| \label{eq:cont_G_trans} .
\end{align}

Suppose now that we have solved the transformed problem and obtained a stopping boundary $c(s) \ge 0$ for $s \le -1$. Then, going back to the original time $r$, it will be optimal to stop if
\begin{equation*}
| W_{s(r)} | \ge c \left( s(r) \right) \iff \sqrt{r_0 + 1} \, |M_r| \ge c \left( s(r) \right) \iff |M_r| \ge \frac{c \left( - \frac{r_0 + 1}{r_0 + r} \right)}{\sqrt{r_0 + 1}} .
\end{equation*}
Therefore, the optimal boundary for the posterior mean process is symmetric, with the part in the positive half-plane given by
\begin{equation} \label{eq:post_mean_boundary}
b_M(r) = \frac{c \left( - \frac{r_0 + 1}{r_0 + r} \right)}{\sqrt{r_0 + 1}}.
\end{equation}
The optimal boundary for the sum process is only symmetric if the prior mean is zero. Let $b_S^+(r)$ and $b_S^-(r)$ denote the upper and lower parts of the boundary, respectively, which together enclose the continuation region for the process. Since $M_r = (m_0 r_0 + S_r)/(r_0 + r)$,
\begin{align*}
  M_r \ge b_M(r) & \iff S_r \ge -m_0 r_0 + (r_0 + r) b_M(r), \\
  M_r \le -b_M(r) & \iff S_r \le -m_0 r_0 - (r_0 + r) b_M(r).
\end{align*}
Hence, $b_S^+(r)$ and $b_S^-(r)$ are given by
\begin{equation} \label{eq:sum_process_boundaries}
  b_S^\pm(r) = -m_0 r_0 \pm \frac{(r_0 + r)c \left( - \frac{r_0 + 1}{r_0 + r} \right)}{\sqrt{r_0 + 1}} .
\end{equation}

Another boundary of interest for us is the one associated with the one-sided p-value process $(p_r)$, which is defined as $p_r \equiv 1 - \Phi \left( S_r / \sqrt{r} \right)$. Its form follows immediately from the definition of $(p_r)$ and equation \eqref{eq:sum_process_boundaries},
\begin{equation} \label{eq:p_value_process_boundary}
  b_p(r) = 1 - \Phi \left( \left( -m_0 r_0 + \frac{(r_0 + r)c \left( - \frac{r_0 + 1}{r_0 + r} \right)}{\sqrt{r_0 + 1}} \right) / \sqrt{r} \right).
\end{equation}
Note the directional change implied by this transformation, i.e., stopping as soon as $S_r \ge b_S^+(r)$ is equivalent to stopping as soon as $p_r \le b_p(r)$. There are two main reasons to study this boundary. Firstly, in the limit $r_0 \to 0$, $b_p(r)$ coincides with a certain boundary for which \citet{Chernoff1981} have obtained asymptotic results (see Section \ref{subsec:asymptotic_results}). Secondly, p-values are often used to support a regulator's decision of whether or not to approve a new medical treatment for marketing. Arguing as in Section \ref{subsec:symmetric_prior}, we obtain
\begin{satz}
For a conjugate normal prior defined by the pair $(m_0,r_0)$, the optimal stopping boundaries for the sum process and the p-value process may be expressed in terms of a single boundary $c(s)$ via equations \eqref{eq:sum_process_boundaries} and \eqref{eq:p_value_process_boundary}. Here, $c(s)$ is the unique solution to the integral equation
\begin{equation} \label{eq:free_boundary_equation_reduced}
  \left(1 + s^{-1} \right) c(s) = \int_s^{-1} \! u^{-2} \, \E_{(s, c(s))} \Big[ | W_u | \, \mathbb{I} \big( | W_u | \ge c(u) \big) \Big] \, \mathrm{d}u , \quad s \le -1,
\end{equation}
in the class of all continuous boundaries.
\end{satz}

\begin{figure}[!htbp] 
  \centering
  \includegraphics[width=0.48 \textwidth]{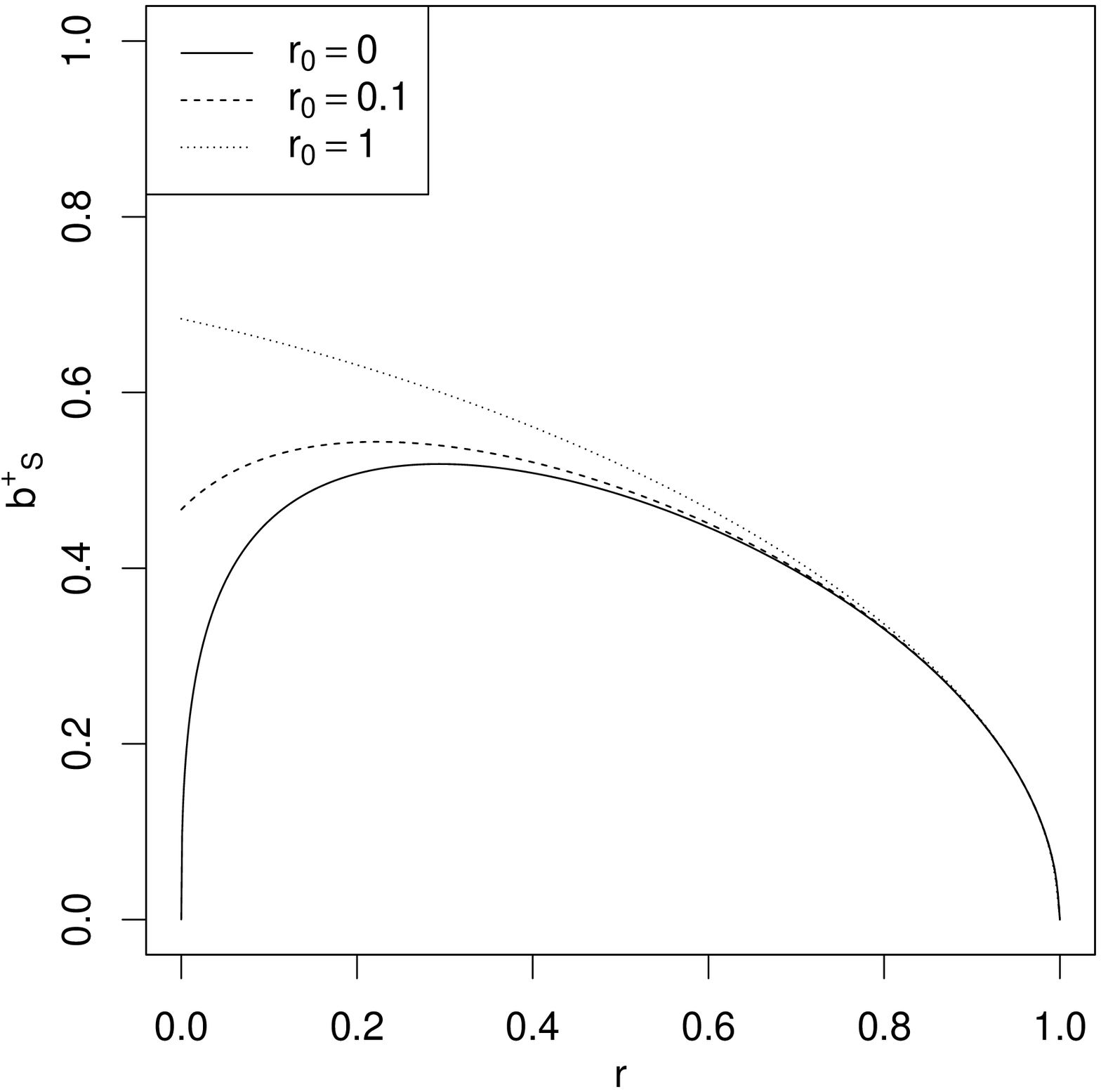} 
  \includegraphics[width=0.48 \textwidth]{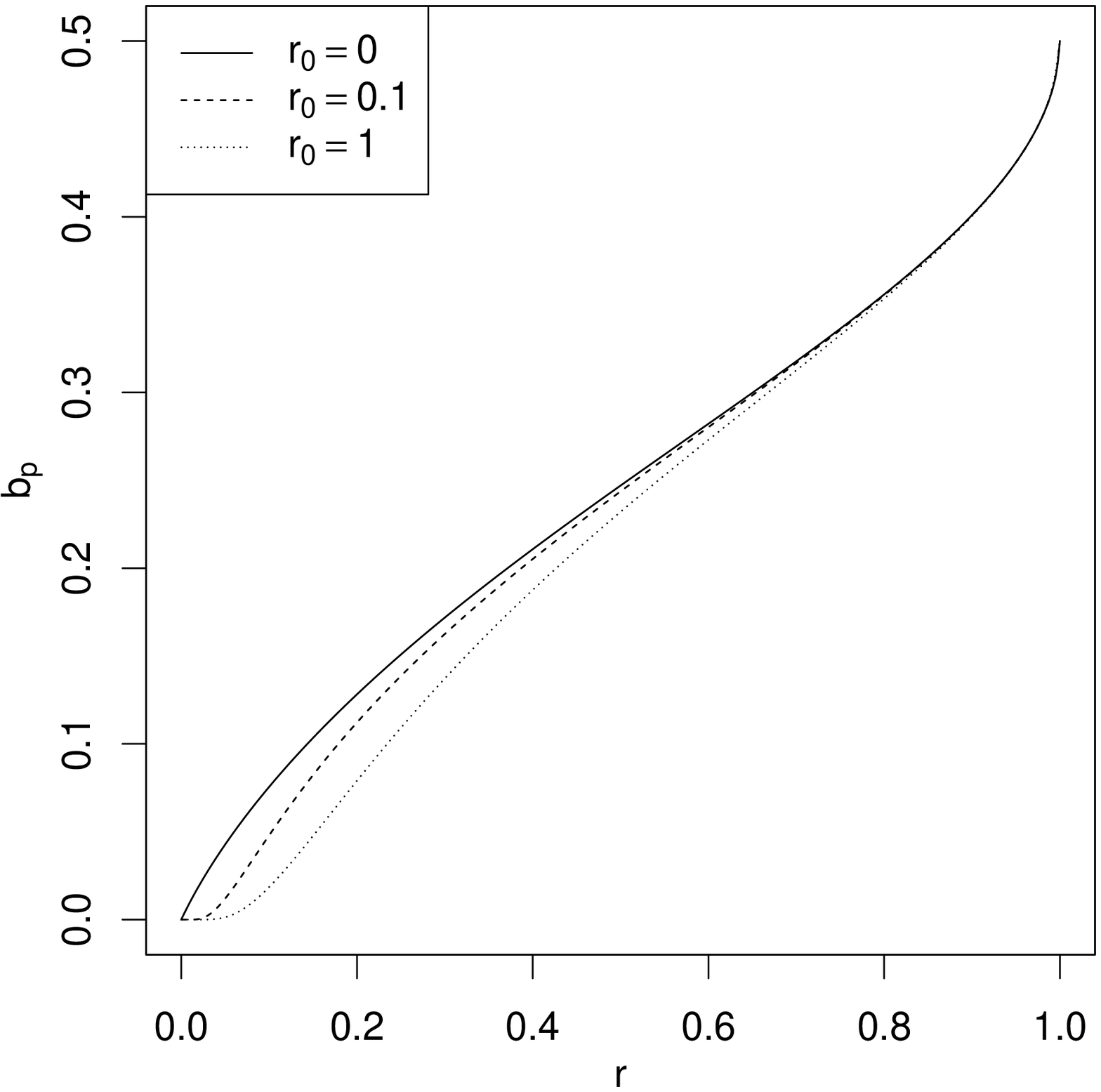} \\
  \caption{$b^+_S(r)$ and $b_p(r)$ for symmetric normal priors.} \label{fig:conjugate_prior} 
\end{figure}

The expectation in the integrand above can be found explicitly in terms of the standard normal density function $\phi$ and distribution function $\Phi$. Such a computation gives
\begin{align*}
  \mathbb{E}_{(s, c(s))} \Big[ | W_u | \, \mathbb{I} \big( | W_u | \ge c(u) \big) \Big] & = \sqrt{u - s} \left( \phi \left( \frac{c(u) - c(s)}{\sqrt{u - s}} \right) + \phi \left( \frac{-c(u) - c(s)}{\sqrt{u - s}} \right) \right) \nonumber \\
  & + c(s) \left( 1 - \Phi \left( \frac{c(u) - c(s)}{\sqrt{u - s}} \right) - \Phi \left( \frac{-c(u) - c(s)}{\sqrt{u - s}} \right) \right) .
\end{align*}
Figure \ref{fig:conjugate_prior} shows $b^+_S(r)$ and $b_p(r)$ for the three symmetric conjugate normal priors defined by $(m_0, r_0) = (0, 0), (0, 0.1), (0, 1)$.

\section{A maximin formulation} \label{sec:maximin}
In this section, we will briefly consider an alternative criterion characterising the optimal stopping time. Instead of maximising \eqref{eq:utility} with respect to a Bayesian prior, we adopt a maximin approach and search for the stopping time which maximises the expected response for the worst case scenario. The problem will be treated in the setting obtained after the standardising transformation described in Section \ref{subsec:standardising_transformation}. It is assumed that the unknown parameter $\delta$ has a known absolute value $\delta_0 > 0$, but that we are unsure about the sign. Motivated by this symmetry, we decide for treatment $A$ if $S_\rho > 0$ and for treatment $B$ otherwise, i.e., we take $D_\rho = \sgn (S_\rho)$. Noting that the multiplicative constant $\delta_0$ appears in front of the expectation for both positive and negative $\delta$, we are then faced with the maximin problem
\begin{equation} \label{eq:freq_two_point_problem}
  \sup_{0 \le \rho \le 1} \min \left( \mathbb{E}^{\delta_0} \big[ (1 - \rho) \sgn (S_\rho) \big], - \mathbb{E}^{-\delta_0} \big[ (1 - \rho) \sgn (S_\rho) \big] \right). 
\end{equation}

\subsection{Reduction to an ordinary stopping problem} \label{subsec:freq_reduc}
The optimal stopping problem \eqref{eq:freq_two_point_problem} does not fall into the setting of worst-case-type optimal stopping problems with unknown drift recently studied by \citet{R,CR}, so that another approach is needed. As in Section \ref{sec:bayes}, we apply Girsanov's transform via the change of measure transformations
\begin{equation*}
  \frac{\mathrm{d} \mathbb{P}|_{\mathcal{F}_r}}{\mathrm{d} \mathbb{P}^{\delta_0}|_{\mathcal{F}_r}} = \exp \left( -\delta_0 S_r + \frac{\delta_0^2}{2} r \right), \quad \frac{\mathrm{d} \mathbb{P}|_{\mathcal{F}_r}}{\mathrm{d} \mathbb{P}^{-\delta_0}|_{\mathcal{F}_r}} = \exp \left( \delta_0 S_r + \frac{\delta_0^2}{2} r \right) .
\end{equation*}
This implies that the two expectations may be written as
\begin{align*}
  E^+ (\rho) & \equiv \E^{{\delta_0}} \left[ (1 - \rho) \sgn (S_\rho) \right] = \E \left[ (1 - \rho) \sgn (S_\rho) \exp \left(\delta_0 S_\rho - \frac{\delta_0^2}{2} \rho \right) \right], \\
  E^- (\rho) & \equiv -\E^{-{\delta_0}} \left[ (1 - \rho) \sgn (S_\rho) \right] = -\E \left[ (1 - \rho) \sgn (S_\rho) \exp \left(-\delta_0 S_\rho - \frac{\delta_0^2}{2} \rho \right) \right],
\end{align*}
where the process $(S_r)$ is a Brownian motion with drift 0 and volatility 1 under the new measure $\mathbb{P}$. Next, define a new expectation $\bar{E} (\rho)$ as
\begin{equation*}
  \bar{E} (\rho) \equiv \frac{E^+ (\rho) + E^- (\rho)}{2} = \E \left[ (1 - \rho) \exp \left( -\frac{\delta_0^2}{2} \rho \right) \sinh \left(\delta_0  |S_\rho| \right) \right] .
\end{equation*}
Since, trivially, $\min \left( E^+ (\rho), E^- (\rho) \right) \le \bar{E} (\rho)$, it follows that
\begin{equation} \label{eq:estimate_min_average}
  \sup_{\rho} \min \left( E^+ (\rho), E^- (\rho) \right) \le \sup_{\rho} \bar{E} (\rho) .
\end{equation}
Now, $\sup_\rho \bar{E} (\rho)$ is a Markovian optimal stopping problem that can be solved using standard techniques, resulting in an optimal stopping time $\rho^*$ that may be expressed in terms of a symmetric boundary for the process $S$. Due to the symmetry of the situation, it is clear here that $E^+ (\rho^*) = E^- (\rho^*)$, implying
\begin{equation*}
  \min \left( E^+ (\rho^*), E^- (\rho^*) \right) = \bar{E}(\rho^*) =\sup_{0 \leq \rho \leq 1} \bar{E} (\rho).
\end{equation*}
Keeping the inequality \eqref{eq:estimate_min_average} in mind, this proves that the stopping time $\rho^*$ is the maximiser for problem \eqref{eq:freq_two_point_problem}, which leads to
\begin{satz}
There exists a maximiser $\rho^*$ for problem \eqref{eq:freq_two_point_problem} and $\rho^*$ is the optimal stopping time for the standard Markovian stopping problem
\begin{equation} \label{eq:bayes-two-point}
  \sup_{0 \leq \rho \leq 1} \E \left[ (1 - \rho) \exp \left(-\frac{\delta_0^2}{2} \rho \right) \sinh \left(\delta_0 |S_\rho| \right) \right].
\end{equation}
\end{satz}

\begin{figure}[!htbp]
  \centering
  \includegraphics[width=0.48\textwidth]{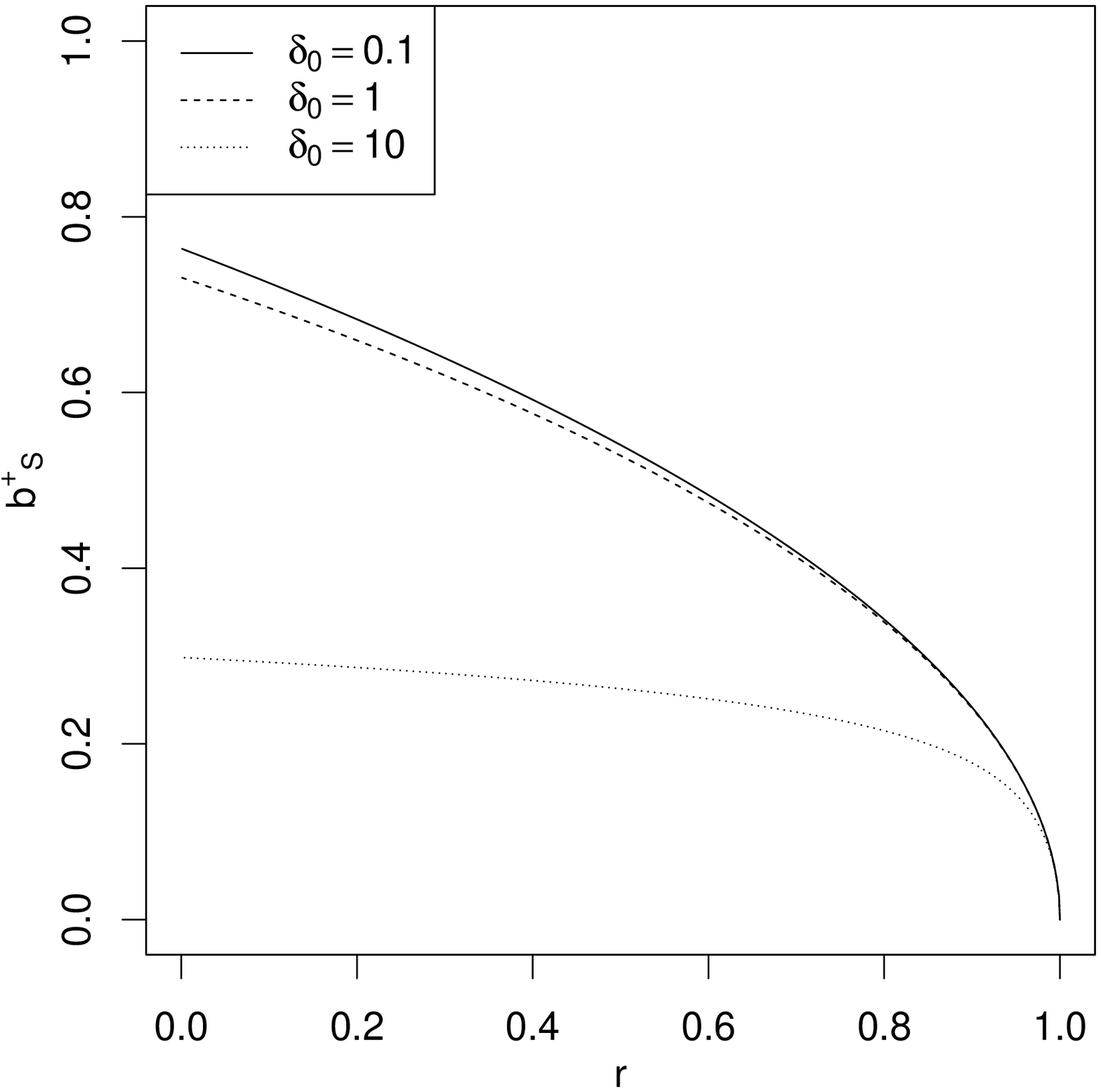}
  \includegraphics[width=0.48\textwidth]{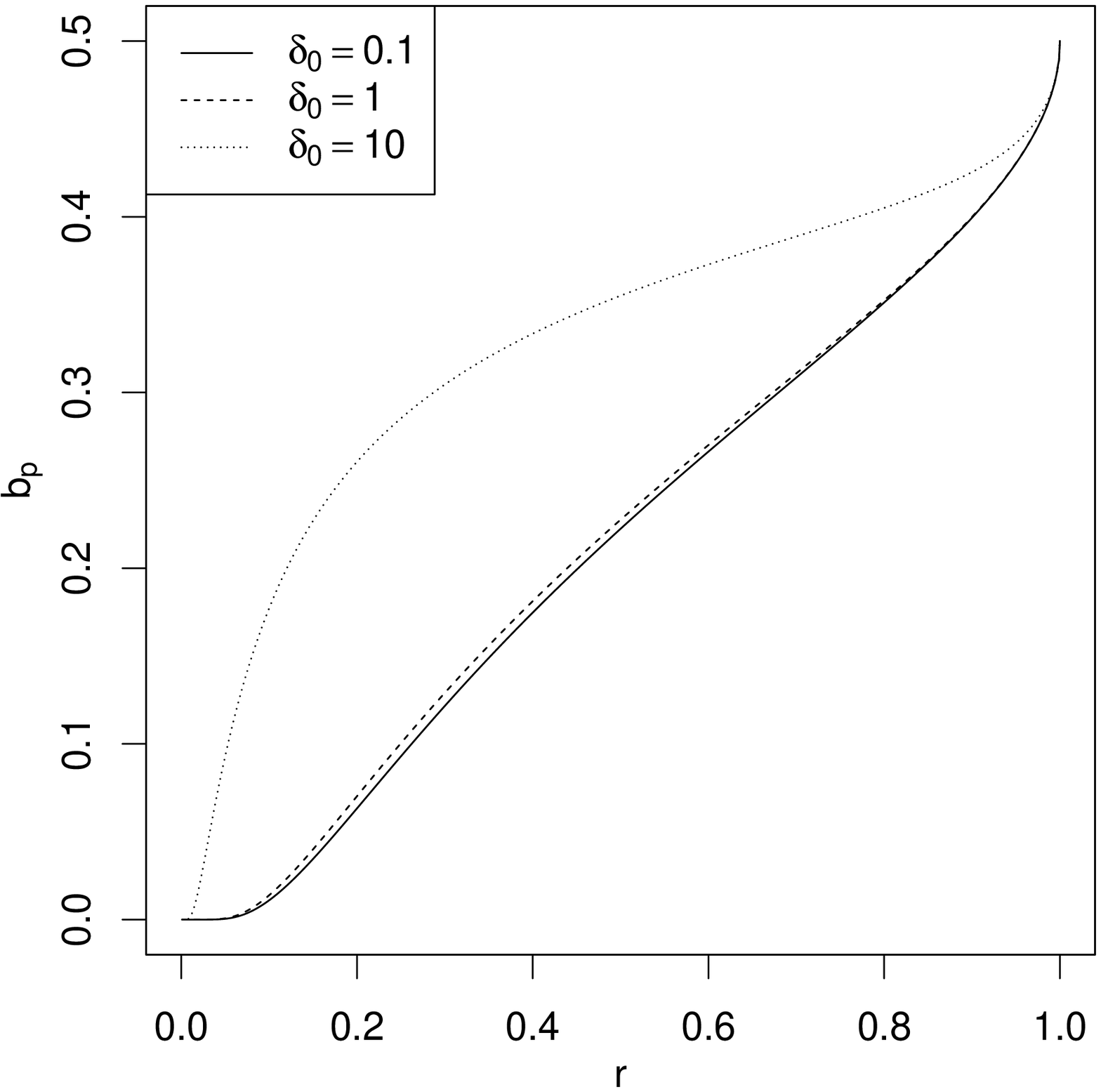} \\
  \caption{$b^+_S(r)$ and $b_p(r)$ for the two-point priors defined by $\delta_0 = 0.1, 1, 10$.} \label{fig:two_point_prior} 
\end{figure}

The auxiliary optimal stopping problem \eqref{eq:bayes-two-point} used for the solution of the maximin problem \eqref{eq:freq_two_point_problem} can be seen as a Bayesian formulation of the decision problem defined by equation \eqref{eq:cont_stopping_problem_transformed} and \eqref{eq:cont_reward_transformed} with a symmetric, two-point prior. Indeed, letting $\xi$ be defined as $\xi(\{\delta_0\})=\xi(\{-\delta_0\})=1/2$, formula \eqref{eq:symmetric_h_function} immediately implies that
\begin{equation*}
  h_\xi (\rho, |S_\rho|) = \delta_0 \exp \left( -\frac{\delta_0^2}{2} \rho \right) \sinh \left( \delta_0 |S_\rho| \right) .
\end{equation*}
Figure \ref{fig:two_point_prior} shows $b^+_S(r)$ and $b_p(r)$ for the three symmetric two-point priors defined by $\delta_0 = 0.1, 1, 10$.

\section{Numerical solution} \label{sec:solving_numerically}
This section describes the numerical method used for solving the optimal stopping problems treated in this paper, which is based on discretising the integral equation for the optimal boundary and then solving it backwards in time. Some alternative approaches to obtain solutions are also briefly discussed. Since numerical issues are not the main focus of this paper, no exhaustive comparison of different procedures based on efficiency or accuracy has been done. The only note we make here is that the integral equation approach is certainly a viable one, and that it provides approximations of reasonable accuracy given the time required to run the algorithm. Moreover, it is straightforward to implement a simple version of the algorithm.

\subsection{Integral equation approach}
Equation \eqref{eq:symmetric_free_boundary_equation_reduced} for the boundary of the standardised problem can be viewed as an instance of a Volterra integral equation of the second kind,
\begin{equation} \label{eq:volterra_integral_equation}
  g(r, b_S(r)) = \int_r^1 \! K \left( u, r, b_S(u), b_S(r) \right) \, \mathrm{d} u, \quad 0 \le r \le 1,
\end{equation}
where the functions $g$ and $K$ are defined as
\begin{align*}
  g(r, y) & \equiv (1 - r) h_\xi (r, y), \\
  K(u, r, z, y) & \equiv \mathbb{E}_{(r, y)} \Big[ h_\xi(u, |S_u|) \, \mathbb{I} \left( | S_u | \ge z \right) \Big] .
\end{align*}

The numerical method uses the trapezoidal rule in order to replace the integral in equation \eqref{eq:volterra_integral_equation} by a sum. Given a grid of $k$ time points $1 = r_1 > r_2 > \ldots > r_k$, letting $b_i = b_S(r_i)$, the discretised version of equation \eqref{eq:volterra_integral_equation} reads
\begin{equation*}
  g(r_i, b_i) \approx \sum_{j = 1}^{i-1} (r_j - r_{j + 1}) \left( \frac{K \left( r_{j + 1}, r_i, b_{j + 1}, b_i \right) + K \left( r_j, r_i, b_j, b_i \right)}{2} \right), \quad 2 \le i \le k .
\end{equation*}
For $i = 2$, the above leads to the algebraic equation
\begin{equation*}
  g(r_2, b_2) = (r_1 - r_2) \left( \frac{K \left( r_2, r_2, b_2, b_2 \right) + K \left( r_1, r_2, b_1, b_2 \right)}{2} \right)
\end{equation*}
for $b_2$, which may be solved using some appropriate numerical procedure. The remaining $b_i$ are then found by solving algebraic equations containing the previously computed values $b_1, \ldots, b_{i - 1}$.

Note that the algebraic equation will include the term $K \left( r_i, r_i, b_i, b_i \right)$ at each step, which can not be computed directly. Such terms are instead computed by replacing $r$ by $r_i$ in $\lim_{u \downarrow r} K(u, r, b_S(u), b_S(r))$. Under the assumption that the function $b_S(r)$ is differentiable, the form of the limit follows from equation \eqref{eq:symmetric_expectation_normal_CDF}:
\begin{align*}
  \lim_{u \downarrow r} K(u, r, b_S(u), b_S(r)) & = \lim_{u \downarrow r} \int_{-\infty}^\infty \! \delta e^{-\delta^2 r / 2} \left\{ e^{\delta b_S(r)} \Phi \left( A_\delta^+ \right) + e^{- \delta b_S(r)} \Phi \left( A_\delta^- \right) \right\} \, \xi(\mathrm{d}\delta) \\
  & = \int_{-\infty}^\infty \! \delta e^{-\delta^2 r / 2} \left\{ e^{\delta b_S(r)} \lim_{u \downarrow r} \Phi \left( A_\delta^+ \right) + e^{- \delta b_S(r)} \lim_{u \downarrow r} \Phi \left( A_\delta^- \right) \right\} \, \xi(\mathrm{d} \delta) \\
  & = \frac{1}{2} \int_{-\infty}^\infty \! \delta e^{-\delta^2 r / 2} e^{\delta b_S(r)} \, \xi(\mathrm{d}\delta) .
\end{align*}

\subsection{Fixed-point equation approach}
Another approach for obtaining a numerical approximation of the optimal boundary is to reformulate equation \eqref{eq:volterra_integral_equation} as a fixed point equation. Since
\begin{align*} 
  & g(r, b_S(r)) = \int_r^1 \! K \left( u, r, b_S(u), b_S(r) \right) \, \mathrm{d} u \iff \\
  & b_S(r) = b_S(r) + \int_r^1 \! K \left( u, r, b_S(u), b_S(r) \right) \, \mathrm{d} u - g(r, b_S(r)),
\end{align*}
if an operator $Q$ is defined as
\begin{equation} \label{eq:fixed_point_operator_definition}
  (Q \, b_S) (r) = b_S(r) + \int_r^1 \! K \left( u, r, b_S(u), b_S(r) \right) \, \mathrm{d} u - g(r, b_S(r)),
\end{equation}
then $b_S$ solves equation \eqref{eq:volterra_integral_equation} if and only if it is a fixed point for $Q$, i.e., if and only if it solves
\begin{equation} \label{eq:fixed_point_equation}
  b_S(r) = (Q \, b_S) (r), \quad 0 \le r \le 1.
\end{equation}
Operationally, the corresponding numerical algorithm may start with a piecewise linear function $b_S^0$ defined on a grid $r_1 > \ldots > r_k$ as $b_{S,i}^0 = b_S^0(r_i) = 0$. A new piecewise linear function $b_S^1$ is then computed by applying the operator $Q$ to $b_S^0$, that is, $b_S^1$ is defined at the grid points as $b_{S,i}^1 = (Q \, b_S^0) (r_i)$. This process is then repeated, producing a sequence of piecewise linear functions $b_S^0, b_S^1, \ldots$. For many of the kernel functions considered in this paper, approximate convergence is obtained after a reasonable number of iterations. However, we found that the trapezoidal method tends to be more computationally efficient.

\subsection{Value function approach}
\citet{Chernoff1986} describe an alternative method for computing numerical approximations for optimal stopping boundaries. First, a finite sequence of time points $r_1 > \ldots > r_k$ is chosen. For simplicity, assume this grid to be uniform, with distance $\Delta_r$ between each point. Let $V$ denote the value function for this discrete problem. Since $V$ corresponds to the optimal decision rule, the backward induction criterion implies that
\begin{equation} \label{eq:discrete_backward_induction}
  V(r_i, y) = \max \left( G(r_i, y), \mathbb{E} \left[ V \left( r_{i-1}, y + Z \sqrt{\Delta_r} \right) \right] \right) , 
\end{equation}
where $Z \sim \mathcal{N}(0, 1)$. Using this equation, it is possible to start at the largest time $r_1$ and compute approximations of $V(r_i, \cdot)$ in steps as $i = 2, \ldots, k$.

The main problem with this approach is that the expectation, taken with respect to a normal distribution, may be too time consuming to evaluate when the time grid is fine enough for the desired accuracy. Hence, another approximation is warranted, and the one suggested by \citet{Chernoff1986} consists of replacing $Z$ with a symmetric Bernoulli distributed random variable $\tilde{Z}$ taking on the values $\pm \sqrt{\Delta_r}$. Equation \eqref{eq:discrete_backward_induction} then becomes
\begin{equation} \label{eq:discrete_backward_induction_bernoulli}
  V(r_i, y) = \max \left( G(r_i, y), \frac{1}{2} \left( V(r_{i-1}, y + \sqrt{\Delta_r}) + V(r_{i-1}, y - \sqrt{\Delta_r}) \right) \right) .
\end{equation}
A possible choice for the grid is
\begin{equation*}
  \left\{ \Big( r_1 - (i - 1) \sqrt{\Delta_r}, \pm j \sqrt{\Delta_r} \Big) : 1 \le i \le k, j \ge 0 \right\}.
\end{equation*}
The continuation and stopping regions can then be found by comparing the computed $V$ with the reward function $G$ in each grid point.\footnote{This is known as the \emph{binomial tree method} in the literature.}

The approach described in this subsection yields the optimal boundary after first computing the value function. In contrast, the method used to produce the plots included in this paper is based on deriving an integral equation for the boundary and then seeking an approximate solution to this equation. In a sense, the latter approach is more direct, since the value function is not explicitly involved. However, sometimes the value function itself is of interest, and a minimal implementation of a solver for the integral equation will not yield the value function without additional computations.

\section{A related asymmetric problem} \label{sec:asymmetric_problem}
In practice, it is unlikely that all patients afflicted by a given disease are available for inclusion in the trial. Even if every patient was asked to participate, some may choose not to. Typically, these patients will still receive some treatment while the trial is in progress. This motivates us to consider a generalised form of Anscombe's problem, and it will be seen that its solution consists of an asymmetric boundary for the optimal stopping region.

As for the original model, let $N$ be the maximum number of patients that may be included in the trial. For each patient included, we will denote by $q$ the number of patients that, for whatever reason, will not participate. Hence, the total number of patients considered is $(q + 1) N$. It is assumed that the non-participants are given the standard treatment $B$ during the trial. The probabilistic model is exactly as before and the objective is now to choose the stopping time $\tau$ and the terminal treatment decision $D_\tau$ so as to maximise the expected utility (i.e., the aggregated treatment response) over all $(q + 1) N$ patients. Since $(q + 1)(N - \tau)$ patients are treated after the trial, the expected utility given $\mu$ for the post-trial patients is $\mu \E^\mu \left[(q + 1) (N - \tau) D_\tau \right]$. In addition, since $q \tau$ patients are treated with $B$ in the trial phase, we have to add an expected utility of $- \mu \E^\mu \left[ q \tau \right]$ and so obtain a total expected utility given $\mu$ of
\begin{equation} \label{eq:utility_asym}
  \mu \E^\mu \left[ (q + 1)(N - \tau) D_\tau - q \tau \right].
\end{equation}

\begin{figure}[!htbp]
  \centering
  \includegraphics[width=0.48 \textwidth]{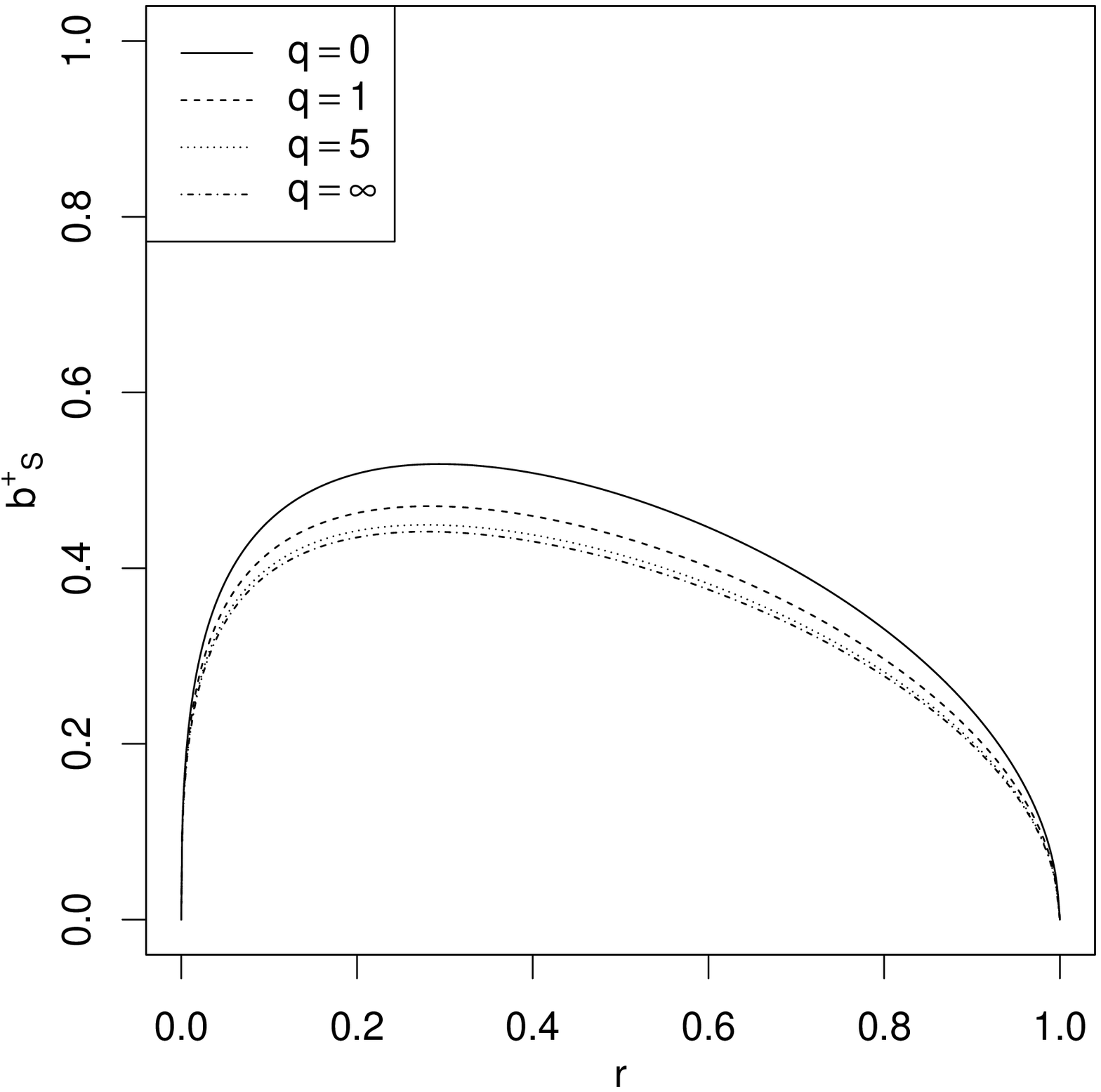} 
  \includegraphics[width=0.48 \textwidth]{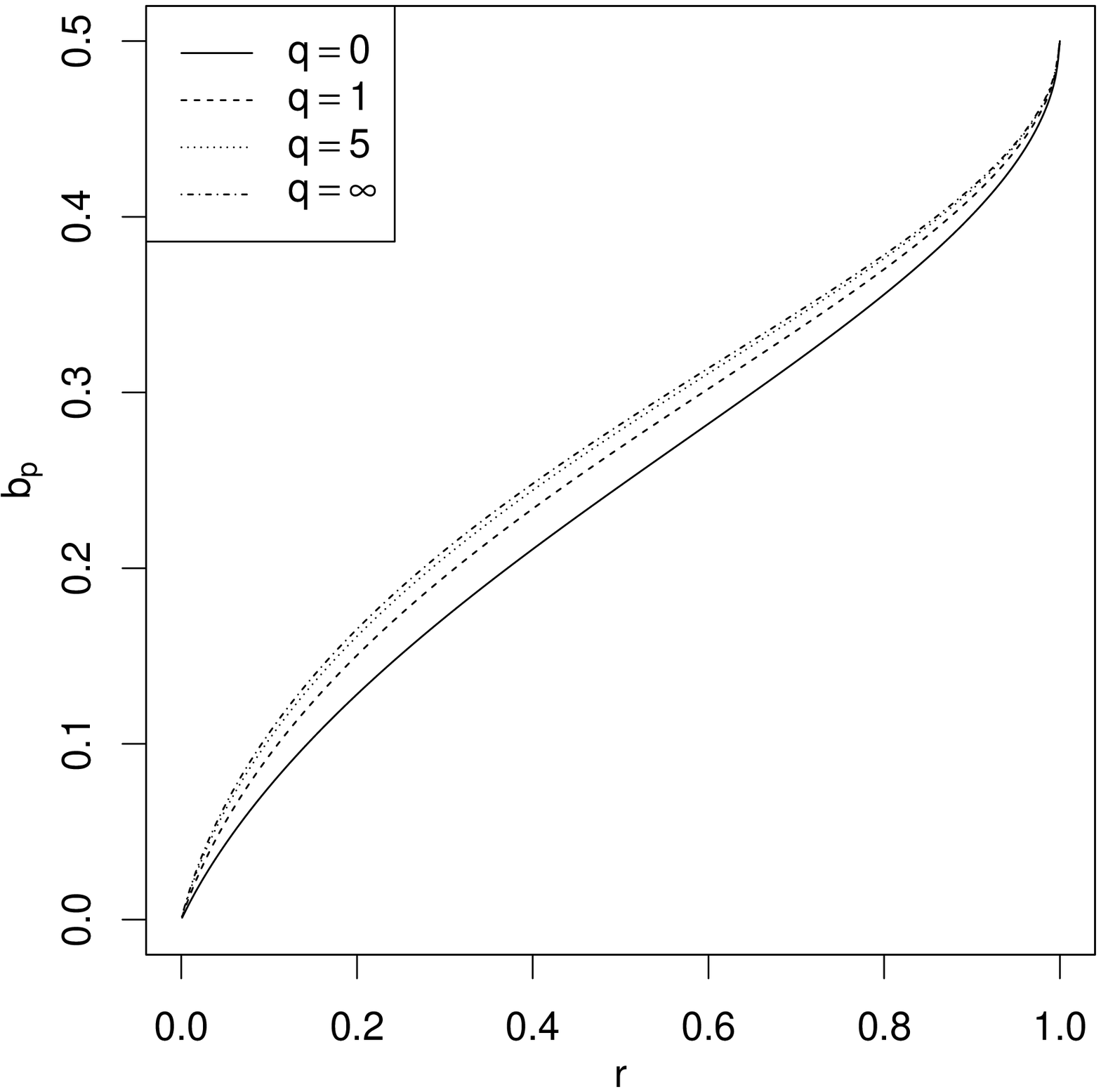} \\
  \caption{$b^+_S(r)$ and $b_p(r)$ for the asymmetric problem. The prior corresponds to the limit case of a conjugate normal one, with $(m_0, r_0) = (0, 0)$.} \label{fig:asymmetric_boundaries} 
\end{figure}

Placing a prior $\nu$ on $\mu$ and arguing as in Section \ref{sec:bayes}, we are faced with the problem of maximising
\begin{equation*}
  \int \! \mu \, \E^\mu \left[ (q + 1)(N - \tau) D_\tau - q \tau \right] \, \nu(\mathrm{d} \mu) = \E \left[ h_\nu (\tau / \sigma^2, \Sigma_\tau / \sigma^2) \Big( (q + 1)(N - \tau) D_\tau - q \tau \Big) \right].
\end{equation*}
As before, $D^*_\tau = \sgn \left( h_\nu(\tau / \sigma^2, \Sigma_\tau / \sigma^2) \right)$ maximises the expectation for each fixed $\tau$, so that the problem is reduced to
\begin{equation*}
  \sup_{0 \le \tau \le N}  \E \left[ \big| h_\nu(\tau / \sigma^2, \Sigma_\tau / \sigma^2) \big| (q + 1)(N - \tau) - h_\nu(\tau / \sigma^2, S_\tau / \sigma^2) q \tau \right].
\end{equation*}
In order to bring the above into a form that more closely resembles the separation of the time and state dependence into two factors (as for the original problem), we add to the above a term with conditional expectation independent of the stopping time. More precisely, as adding the martingale $\left( q N h_\nu(t / \sigma^2, \Sigma_t / \sigma^2) \right)_{t \in [0,N]}$ does not change the optimal stopping time, the problem is equivalent to that of maximising
\begin{align*}
  & \E \left[ \big| h_\nu(\tau / \sigma^2, \Sigma_\tau / \sigma^2) \big| (q + 1)(N - \tau) - h_\nu(\tau / \sigma^2, \Sigma_\tau / \sigma^2) q \tau + q N h_\nu(\tau / \sigma^2, \Sigma_\tau / \sigma^2) \right] = \\
  & \E \left[ (N - \tau) \left( \big| h_\nu(\tau / \sigma^2, \Sigma_\tau / \sigma^2) \big| + 2 q h_\nu(\tau / \sigma^2, \Sigma_\tau / \sigma^2)^+ \right) \right],
\end{align*}
where $h_\nu^+$ denotes the positive part of $h_\nu$. We have now proved
\begin{satz}
For the problem of maximising the expected value of \eqref{eq:utility_asym} with respect to a prior $\nu$ on $\mu$, the optimal decision variable is given by $D^*_\tau= \sgn \left( h_\nu(\tau / \sigma^2, \Sigma_\tau / \sigma^2) \right)$ and the optimal stopping time $\tau^*$ solves the Markovian optimal stopping problem
\begin{equation} \label{eq:problem_asym}
  \sup_{0 \leq \tau \leq N} \E \left[ (N - \tau) \left( \big| h_\nu(\tau / \sigma^2, \Sigma_\tau / \sigma^2) \big| + 2 q h_\nu(\tau / \sigma^2, \Sigma_\tau / \sigma^2)^+ \right) \right].
\end{equation}
\end{satz}
Application of the standardising time-space transformation described in Section \ref{subsec:standardising_transformation} now leads to the problem
\begin{equation} \label{eq:problem_asym_standardised}
  \sup_{0 \leq \rho \leq 1} \E \left[ (1 - \rho) \left( \big| h_\xi (\rho, S_\rho) \big| + 2 q h_\xi(\rho, S_\rho)^+ \right) \right],
\end{equation}
where, as before, $\xi$ is the prior on $\delta$ induced by the relation $\delta = \mu \frac{\sqrt{N}}{\sigma}$. As expected, when $q = 0$, problem \eqref{eq:problem_asym_standardised} is reduced to the original symmetric problem defined by equations \eqref{eq:cont_stopping_problem_transformed} and \eqref{eq:cont_reward_transformed}. It is interesting to consider the effect upon the solution when $q \to \infty$, corresponding to the case in which a large number of patients will be given the standard treatment in parallel with the trial. For example, this would be the case if the targeted disease is relatively common. Since problem \eqref{eq:problem_asym_standardised} is equivalent to the one obtained when dividing through by $2 q$, we obtain the following (non-degenerated) problem in the limit $q \to \infty$:
\begin{align} \label{eq:asymp_q}
  \sup_{0 \leq \rho \leq 1} \E \left[ (1 - \rho) h_\xi(\rho, S_\rho)^+ \right].
\end{align}
Figure \ref{fig:asymmetric_boundaries} shows $b^+_S(r)$ and $b_p(r)$ when an uninformative normal conjugate prior is used for $q = 0, 1, 5, \infty$.

\subsection{Asymptotic results when $r_0$ and $r$ are small} \label{subsec:asymptotic_results}
Consider the case of a normal conjugate prior for $\delta$. Recall that the time parameter $r$ was defined as the ratio $r = n / N$ and that $r_0$ was defined as $n_0 / N$, where $n_0$ is the number of virtual observations in the prior. In many cases, it seems reasonable to assume that the total population size $N$ is very large compared to $n_0$, thus making $r_0$ very small. In the beginning of the trial, the optimal strategy is determined by the shape of the boundary when $r$ is small. Hence, the asymptotic behaviour of the boundaries when both $r_0$ and $r$ are small is of particular interest for the implementation of the optimal procedure at the beginning of the trial.

Optimal boundary asymptotics for Anscombe's problem have previously been studied by \citet{Chernoff1981}. In the remaining part of this section, we will describe how their derivation can be modified so as to also cover the asymmetric problem with $q > 0$. Further, it will be shown how this result can be used to derive an approximate expression for the optimal boundary of the p-value process. For the asymmetric problem, the standardised optimal stopping problem is
\begin{align*}
  \hat{V} (s, y) & = \sup_{s \le \zeta \le -1} \E_{(s, y)} \left[ \hat{G} (\zeta, W_\zeta) \right], \\
  \hat{G} (s, y) & = \left(1 + s^{-1} \right) \left( |y| + 2 q y^+ \right)  .
\end{align*}
By equation \eqref{eq:time_transformation}, small values of $r_0$ and $r$ correspond to large negative values for $s$. As $s \to -\infty$, we obtain the approximation
\begin{equation*}
  \hat{G} (s, y) = \left(1 + s^{-1} \right) \left( |y| + 2 q y^+ \right) \approx \hat{G}_{-\infty} (y) \equiv |y| + 2 q y^+ .
\end{equation*}
This suggests that the value function $\hat{V}$ solving the problem may be approximated (as $s \to -\infty$) with a solution to the equation $\left( \frac{\partial}{\partial s} + \frac{1}{2} \frac{\partial^2}{\partial^2 y} \right) \hat{V} = 0$ that is approximately equal to $\hat{G}_{-\infty} (y)$. Setting $z = y (-s)^{-1/2}$, such a function is given by
\begin{align*}
  \E \left[ \hat{G}_{-\infty} (W_0) \mid W_s = y \right] & = 2 (-s)^{1/2} \psi_q ( z ), \\
  \psi_q ( z ) & \equiv \phi (z) (1 + q) + z \left(\Phi (z)(1 + q) - 1/2 \right) .
\end{align*}
Note that $\E \left[ \hat{G}_{-\infty} (W_0) \mid W_s \right]$ is a martingale when considered as a process in time $s$. Hence, the optimal boundary may be found by solving the problem obtained by replacing $\hat{G}$ with the modified reward function $\hat{G}_0 = \hat{G} (s, y) - 2 (-s)^{1/2} \psi_q ( z )$. In this modified reward, the major part of $\hat{G} (s, y)$ has been cancelled by the martingale. This facilitates the matching of terms in the asymptotic expansion of the boundary. The remaining steps of the derivation leading to the result below closely follows \citet{Chernoff1981} and may be found in Appendix \ref{sec:asymptotic_expansion}.
\begin{satz} \label{thm:asymptotic_expansion}
  Let $c^+_q (s)$ be the upper part of the optimal boundary for the standardised version of the asymmetric problem with a normal conjugate prior. Then  
  \begin{equation}
    c^+_q (s) \sim \sqrt{-s} \, \Phi^{-1} \left( 1 + \left( \frac{1 + 2q}{1 + q} \right) s^{-1} \right), \quad s \to -\infty . \label{eq:cq_asymptotic_exp}
  \end{equation}
\end{satz}
The p-value process boundary is defined in terms of $c^+_q (s)$ according to
\begin{equation*}  
  b_p(r, q) = 1 - \Phi \left( \left( -m_0 r_0 + \frac{(r_0 + r) c^+_q \left( - \frac{r_0 + 1}{r_0 + r} \right)}{\sqrt{r_0 + 1}} \right) / \sqrt{r} \right).
\end{equation*}
When both $r_0$ and $r$ are small, the result \eqref{eq:cq_asymptotic_exp} therefore leads to the approximation
\begin{equation} \label{eq:p_value_boundary_approx} 
  b_p(r, q) \approx 1 - \Phi \left( \frac{ -m_0 r_0 + \sqrt{r_0 + r} \Phi^{-1} \left( 1 - \left( \frac{1 + 2q}{1 + q} \right) \left( \frac{r_0 + r}{r_0 + 1} \right) \right) }{\sqrt{r}} \right).
\end{equation}
The limit case of an uninformative prior corresponds to $r_0 = 0$, which gives the simpler form
\begin{equation} \label{eq:p_value_boundary_uninformative_approx}
  b_p(r, q) \approx 1 - \Phi \left( \Phi^{-1} \left( 1 - \left( \frac{1 + 2q}{1 + q} \right) r \right) \right) = \left( \frac{1 + 2q}{1 + q} \right) r .
\end{equation}

\section{Model with a random number of patients} \label{sec:random_nb}
One questionable assumption that has been made up to this point is that the patient horizon $N$ is known. In this section, we generalize the model by allowing $N$ to be a $(0,\infty)$-valued random variable with a finite mean. For example, $N$ could be interpreted as a time point at which a new (and highly competitive) drug arrives. At time $N$, we assume the testing to stop immediately. In analogy with equation \eqref{eq:utility}, for a given $\mu$, the expected utility reads as
\begin{equation*}
  \mu \E^\mu \Big[ (N - \tau) D_\tau \, \mathbb{I} \{ \tau \leq N \} \Big] = \mu \E^\mu \Big[ (N - \tau)^+ D_\tau \Big].
\end{equation*}

It is assumed that the random variable $N$ is independent of the underlying process for each given $\mu$, so that we are faced with the expected utility
\begin{equation*}
  \mu \E^\mu \Big[ (N - \tau)^+ D_\tau \Big] = \mu \E^\mu \Big[ f_\mu (\tau) D_\tau \Big], \quad f_\mu(t) \equiv \E^\mu \Big[ (N - t)^+ \Big].
\end{equation*}
By also making the further assumption that $N$ is independent of the drift $\mu$, it follows that the value of $f_\mu(t)$ is the same for all $\mu$. This common value will be denoted by $f(t)$. Following the derivation of Section \ref{sec:bayes}, the independence assumptions for $N$ now leads to
\begin{satz}
With an independent random time horizon $N$ and a prior $\nu$ on $\mu$, the optimal decision variable is given by $D^*_\tau = \sgn \left( h_\nu(\tau / \sigma^2, \Sigma_\tau / \sigma^2) \right)$ and the optimal stopping time $\tau^*$ solves the Markovian optimal stopping problem
\begin{equation} \label{eq:random_N_stopping_problem}
  \sup_{0\leq \tau<\infty} \E \left[ f(\tau) \big| h_\nu(\tau / \sigma^2, \Sigma_\tau / \sigma^2) \big| \right].
\end{equation}
\end{satz}
As done in Section \ref{subsec:standardising_transformation}, a standardising transformation can be performed also in the case of a random $N$. For this, $\delta$ is defined using the expected value of $N$ according to $\delta = \mu \sqrt{\E[N]} / \sigma$, with the measure for $\delta$ denoted by $\xi$ as before. The new time parameter $r$ is defined as the fraction of the expected total number of patients, $r = t / \E[N]$ (giving $\rho = \tau / \E[N]$). This leads to the standardised optimal stopping problem
\begin{equation*}
  \sup_{0 \leq \rho < \infty} \E \left[ \tilde{f}(\rho) \big| h_\xi(\rho, S_\rho) \big| \right], \quad \tilde{f}(r) \equiv \frac{f(r \E [N])}{\E [N]},
\end{equation*}
in which $(S_r)$ is a Brownian motion with zero drift and unit volatility.

\subsection{Some general properties of $\tilde{f}$ and some examples}
It is immediately clear that $\tilde{f}$ is non-increasing in $r$, $\tilde{f}(0) = 1$ and $\lim_{r \to \infty} \tilde{f}(r) = 0$. Under the assumption that $N$ has a density $\pi_N$, we also have
\begin{equation}
  f'(t) = \int_0^\infty \! \frac{\partial}{\partial t} \left( (u - t)^+ \right) \pi_N (u) \, \mathrm{d} u = \int_0^\infty \! (-1) \mathbb{I} \left( t < u  \right) \pi_N (u) \, \mathrm{d} u = - \mathbb{P} \left( N > t \right) . \label{eq:h_derivative}
\end{equation}
Equation \eqref{eq:h_derivative} implies that $\tilde{f}'(r) = f'(r \E[N]) = - \mathbb{P} \left( N > r \E[N] \right)$ and, in particular, $\tilde{f}'(0) = -1$. Since $f$ is an integral of convex functions, it is also convex. Conversely, given a convex, non-increasing function $f$ on $(0,\infty)$, a distribution for $N$ may be found that generates $f$ via the identity $f'(t)= -\mathbb{P} \left( N > t \right)$.

\begin{bsp}\label{ex:exp}
Suppose $N \sim \text{Exp}(\lambda)$. Making use of the memoryless property of the exponential distribution and that $\E [N] = 1 / \lambda$,
\begin{align*}
  f(t) & = \E \left[ (N - t) \mid N > t \right] \mathbb{P} \left( N > t \right) = e^{-\lambda t} / \lambda , \\
  \tilde{f}(r) & = \frac{f(r \E [N])}{\E [N]} = \lambda f(r/\lambda) = e^{-r} .
\end{align*}
\end{bsp}

\begin{bsp}\label{bsp:lomax}
Suppose $N \sim \text{Lomax}(\lambda, \omega)$ \citep{Lomax1954}, i.e.,
\begin{equation*}
F_N (u) = 1 - \left( \frac{\lambda}{\lambda + u} \right)^\omega , \quad u \ge 0, \quad \mathbb{E}[N] = \frac{\lambda}{\omega - 1} ,
\end{equation*}
where $\lambda > 0$ is a scale parameter, $\omega > 1$ is a shape parameter and $F_N$ is the distribution function of $N$. This distribution can be viewed as a Pareto Type I distribution, shifted so that its support begins at zero. It has the property that $N - t | N > t \sim \text{Lomax}(\lambda + t, \omega)$, implying that
\begin{equation*}
\mathbb{E} [N - t \mid N > t] = \frac{\lambda + t}{\omega - 1}, \quad \mathbb{P} (N > t) = \left( \frac{\lambda}{\lambda + t} \right)^\omega .
\end{equation*}
It follows that
\begin{align*}
  f(t) & = \E \left[ N - t \mid N > t \right] \mathbb{P} \left( N > t \right) = \left( \frac{\lambda^\omega}{\omega - 1} \right) \left( \lambda + t \right)^{1 - \omega} , \\
  \tilde{f}(r) & = \frac{f(r \E [N])}{\E [N]} = \left( \frac{\omega - 1}{\lambda} \right) \left( \frac{\lambda^\omega}{\omega - 1} \right) \left( \lambda + r \frac{\lambda}{\omega - 1} \right)^{1 - \omega} = \left( 1 + \frac{r}{\omega - 1} \right)^{1 - \omega} .
\end{align*}
\end{bsp}

\subsection{The maximin problem with an exponential number of patients} \label{subsec:maximin_exp}
Considering the maximin problem from Section \ref{sec:maximin} with $N \sim \text{Exp}(\lambda)$, we are --- keeping Example \ref{ex:exp} in mind --- faced with the optimal stopping problem
\begin{align*} \label{eq:maximin_exponential}
  \sup_{0 \leq \rho < \infty} \E \left[ \tilde{f} (\rho) e^{-\frac{\delta_0^2}{2} \rho} \sinh \left(\delta_0 |S_\rho| \right) \right]=\sup_{0 \leq \rho < \infty} \E \left[e^{-\left( \frac{\delta_0^2}{2} + 1 \right) \rho} \sinh \left(\delta_0 |S_\rho| \right) \right].
\end{align*}
This is a problem with infinite time horizon and discounting that can be solved with standard techniques such as the change of measure technique (see \cite{BL,CI2}). The differential equation
\[\frac{1}{2}y''= \left( \frac{\delta_0^2}{2}+1 \right)y\]
has the symmetric solution
\begin{equation*}
  y(x) = \cosh(C x), \quad C \equiv \sqrt{\delta_0^2+2} .
\end{equation*}
This implies (see \citet[Theorem 4]{BL}) that the optimal stopping time is $\rho_1^* = \inf \{r \geq 0: |S_r| \geq x_1^* \}$, where $x_1^*$ is the unique maximum point of $\sinh(\delta_0 x) / \cosh(Cx)$, i.e., the unique positive solution of the algebraic equation
\[\delta_0 \cosh(\delta_0 x) \cosh(Cx)- C \sinh(\delta_0 x) \sinh(Cx) =0 .\]
Hence, the optimal boundary for the sum process is constant and completely determined by the value of $x_1^*$. Note also that the expectation of $\rho^*$ is $\E [\rho^*] = (x_1^*)^2$ \citep{Borodin}.

Next, consider the asymmetric problem obtained by letting $q \to \infty$. We are then faced with the optimal stopping problem
\begin{equation*}
  \sup_{0 \leq \rho < \infty} \E \left[ e^ { -\left( \frac{\delta_0^2}{2} + 1 \right) \rho } \sinh \left(\delta_0 S_\rho \right)^+ \right].
\end{equation*}
This problem is of one-sided type and we obtain --- using \citet[Theorem 2]{BL}  --- that the optimal stopping time is given by $\rho_2^* = \inf \{r \geq 0: S_r \geq x_2^* \}$, where $x_2^*$ is the unique solution to
\begin{equation*}
  \delta_0 \cosh(\delta_0 x)\exp(C x) - C \sinh(\delta_0 x) \exp(C x) = 0 .
\end{equation*}
Solving this equation gives
\begin{equation*}
  x_2^* = \frac{\ln \left( \frac{\sqrt{\delta_0^2 + 2} + \delta_0}{\sqrt{\delta_0^2 + 2} - \delta_0} \right)}{2 \delta_0} = \frac{\ln \left( \sqrt{\frac{\delta_0^2}{2} + 1} + \frac{\delta_0}{\sqrt{2}} \right)}{\delta_0}.
\end{equation*}

\subsection{On the normal prior with a random time horizon}
One advantage of the model extension discussed in this section is that it respects the additional structure given by the normal distribution. Indeed, proceeding as in Section \ref{sec:normal_free_bound}, it can be seen that the standardised optimal stopping problem for the Brownian motion $(W_s)$ analogous to equations \eqref{eq:cont_stopping_problem_trans} and \eqref{eq:cont_G_trans} is given by
\begin{align*}
  \hat{V} (s, y) & = \sup_{s \le \zeta < 0} \mathbb{E}_{(s, y)} \left[ \tilde{G} (\zeta, W_\zeta) \right],  \\
  \hat{G} (s, y) & = \tilde{f} \left( -r_0 - \frac{r_0 + 1}{s} \right) |y| .
\end{align*}
It is interesting to discuss the case in which $N$ has a Lomax distribution. By Example \ref{bsp:lomax},
\begin{equation*}
  \tilde{f} \left( -r_0 - \frac{r_0 + 1}{s} \right) = \left( 1 + \frac{ -r_0 - \frac{r_0 + 1}{s}}{\omega - 1} \right)^{1 - \omega} = \left(\frac{\omega - 1 - r_0 - \frac{r_0 + 1}{s}}{\omega - 1} \right)^{1 - \omega} .
\end{equation*}
Recall that, in the case of a known $N$, $r_0$ was defined as $r_0 = n_0 / N$. For a Lomax-distributed $N$, the definition is instead $r_0 = n_0 / \mathbb{E}[N] = n_0 (\omega - 1) / \lambda$. The special case of $\lambda = n_0$ will be considered here, since this assumption allows us to obtain an explicit functional form for the optimal stopping boundary of the standardised problem. Note that this assumption implies that $\omega = r_0 + 1$. Using the expression for $\tilde{f}$ above, and noting that the optimal boundary is not affected by a constant depending only on the parameters, the optimal stopping problem to solve is
\begin{equation}
  \hat{V} (s, y) = \sup_{s \le \zeta < 0} \mathbb{E}_{(s, y)} \Big[ (-\zeta)^{r_0} |W_\zeta| \Big] = \sup_{0 \le \zeta < -s} \mathbb{E} \Big[ \left( -(s + \zeta) \right)^{r_0} |y + W_\zeta| \Big] . \label{eq:Lomax_standardised_problem}
\end{equation}
The method now consists of transforming \eqref{eq:Lomax_standardised_problem} into a solvable problem. Our approach essentially follows the time-change method described by \citet[Section IV{.}10]{ps}.  Rewriting the argument of the expectation in \eqref{eq:Lomax_standardised_problem} as
\begin{equation*}
  \left( -(s + \zeta) \right)^{r_0} |y + W_\zeta| = (-s)^{r_0} \left( 1 - \zeta / (-s) \right)^{r_0} \sqrt{-s} \left| \frac{y}{\sqrt{-s}} + \frac{1}{\sqrt{-s}} W_{(-s) (\zeta / (-s))} \right|,
\end{equation*}
setting $\tilde{\zeta} = \zeta / (-s)$ and noting that, by Brownian scaling, $\tilde{W}_{\tilde{s}} = (1 / \sqrt{-s}) W_{(-s) \tilde{s} }$ is a standard Brownian motion, \eqref{eq:Lomax_standardised_problem} becomes
\begin{equation*}
\hat{V} \left( \tilde{w} \right) = \sup_{0 \le \tilde{\zeta} < 1} \mathbb{E}_{\tilde{w}} \left[ \left( 1 - \tilde{\zeta} \right)^{r_0} | \tilde{W}_{\tilde{\zeta}} | \right] .
\end{equation*}
In this problem, the process $(\tilde{W}_{\tilde{s}})$ starts at $\tilde{w} = \frac{y}{\sqrt{-s}}$. The next step consists of performing the transformation
\begin{equation*}
  \hat{Y}_{\tilde{s}} = \frac{1}{\sqrt{1 - \tilde{s}}} \tilde{W}_{\tilde{s}}, \quad \tilde{s}(\hat{s}) = 1 - e^{-2 \hat{s}}, \quad \hat{W}_{\hat{s}} = \hat{Y}_{ \tilde{s}(\hat{s}) } .
\end{equation*}
This makes $(\hat{W}_{\hat{s}})$ an Ornstein-Uhlenbeck process with generator 
$\mathbb{L}_{\hat{W}} = \hat{w} \frac{\partial}{\partial \hat{w}} + \frac{\partial^2}{\partial \hat{w}^2}$ and the stopping problem becomes
\begin{equation*}
  \hat{V}(\hat{w}) = \sup_{0 \le \hat{\zeta} < \infty} \mathbb{E}_{\hat{w}} \left[ e^{-(2 r_0 + 1)\hat{\zeta}} |\hat{W}_{\hat{\zeta}}| \right] , \text{ where } \hat{w} = \tilde{w} = \frac{y}{\sqrt{-s}} .
\end{equation*}
Let $\mathcal{M} (\cdot, \cdot, \cdot)$ denote the Kummer confluent hypergeometric function. It may then be shown \citep[Theorem 4]{BL} that the optimal stopping time in this transformed problem is $\hat{\zeta}^*=\inf\{\hat{s} \geq 0: |\hat{W}_{\hat{s}}|\geq \hat{w}^* (r_0) \}$, where $\hat{w}^*(r_0)$ is the unique solution to the equation
\begin{equation} \label{eq:optimal_point_equation}
  2 (r_0 + 1) \frac{\mathcal{M} \left( r_0 + 2, 3/2, \hat{w}^2 / 2 \right)}{\mathcal{M} \left( r_0 + 1, 1/2, \hat{w}^2 / 2 \right)} \hat{w}^2 = 1 + \hat{w}^2 .
\end{equation}
Transforming back to the original variable pair $(s, W_s)$, it follows that the optimal boundary for problem \eqref{eq:Lomax_standardised_problem} has the form $c(s) = \hat{w}^*(r_0) \sqrt{-s}$.

\section{Discussion} \label{sec:discussion}

\subsection{On the use of Girsanov's transformation}
An early application of Girsanov's transformation is central to our treatment of Anscombe's problem. The transformation allows for treating different priors in a common framework: that of an optimal stopping problem for a Brownian motion with zero drift. In particular, we are by this route led to the observation that the optimal stopping boundary for any normal, conjugate prior can be obtained by solving a single optimal stopping problem. That this is possible was noted by \citet{Chernoff1981}.

\subsection{Model applications} \label{subseq:model_applications}
In Anscombe's formulation of the treatment selection problem, a number of simplifying assumptions are intentionally made. Trial costs are ignored, the production costs of the treatments once in usage are assumed to be the same and there are no costs involved when switching from one treatment regime to another. Although it may be argued that the basic model is not detailed enough to be directly applicable to practical trial design problems, it can be extended for increased realism. For example, a recent contribution by \citet{Pertile2014} makes use of Chernoff's free-boundary approach to solve a model taking treatment costs into account. The primary quantity of interest in their context is not the incremental effect of $A$ relative to $B$, but the net incremental monetary benefit, which is obtained as the difference between the incremental effect (in monetary units) and the incremental cost (on a per-patient level). Such an extended model is motivated by the increasing focus on not only obtaining effective medications, but cost-effective ones. Their analysis shows the flexibility of Chernoff's approach and indicates an applicability that far transcends Anscombe's streamlined problem.

It is important to note that the perspective of the trial designer has an impact on what kind of design that will be considered optimal. For publicly sponsored trials it seems reasonable to assume that the main goal is to maximise the total public health. Anscombe's model, possibly extended with trial costs, would then seem appropriate. If the sponsor is a pharmaceutical company, the goal is instead to find a design maximising expected profit. Situations in which the profit is directly proportional to the aggregated health benefit for the population, while certainly conceivable, would hardly be the most common in practice.

In commercial drug development, a company with a new candidate drug plans a trial, executes it and hands over the resulting data to a regulatory authority. Based on the trial design and the resulting data, the regulator then either grants market approval or rejects the new medication. In some cases, it might also ask the company to provide data from an extended study before a final decision is made. The form of the decision rule used by the regulator determines which medicines are approved for marketing and therefore impacts public health. This raises the question of which particular rule to use. The canonical procedure used by the FDA in the US has been to require significant statistical results in two independent studies \citep{FDA98}. For historical reasons, a significance level of 5\% is often used\footnote{This particular value for the level goes back to a suggestion made by \citet{Fisher1946}.}, regardless of the prior information available before the trial, and also without taking the sample size or the total population size into account. 

The results obtained when analysing Anscombe's model leads to an argument for using a regulatory rule which is different from the classical one. In the interest of simplicity, let us assume that the regulator's decision is binary, i.e., either the medicine is approved or it is not. Being presented with the trial data, the regulator can then map this data to a point which will lie in the continuation or stopping region corresponding to the optimal boundary for Anscombe's model. If the point ends up in the stopping region, then it would clearly be suboptimal from a public health perspective to reject the new treatment. For if we imagine that the authority takes control of the trial procedure and continues optimally, then it will realise that optimality dictates immediate stopping (and distribution of the new medication). It is not as clear what the regulator should do if the trial data yields a point in the continuation region. To continue would be optimal if it could be done sequentially, but now a regulatory decision must be taken directly. Sometimes it will be optimal to approve, and sometimes to reject. Hence, while the argument does not lead to a complete specification of a replacement for the classical rule, it does provide a motivation for using the optimal nominal p-value curve as a source of lower bounds for significance level tests.

We will now use the approximation $b_p(r) = b_p(r, q = 0) \approx r$ derived in Section \ref{subsec:asymptotic_results}, which holds in limit case of an uninformative prior, to give a detailed example of the argument in the preceding paragraph. Consider two different scenarios faced by a regulator:
\begin{enumerate}
\item Common disease and large trial: $N_1 = 10^8$ and $n_1 = 10^3$, giving $r_1 = \frac{n_1}{N_1} = 10^{-5}$.
\item Rare disease and small trial: $N_2 = 10^5$ and $n_2 = 10^2$, giving $r_2 = \frac{n_2}{N_2} = 10^{-3}$.
\end{enumerate}
The rule $b_p(r)$ is compared with a model $b^c_p(r)$ of the classical procedure in which statistical significance at level $\alpha$ is required in two independent trials if a new treatment is to be accepted. Since we only consider one trial here, the independence assumption implies that $b^c_p(r) = \alpha^2$. It is appropriate here to use one-sided significance levels and, following tradition, we choose $\alpha = 0.025$, giving $b^c_p(r) = 0.000625$. Note that there are no sequential decisions by the regulator here, even though $b_p(r)$ was derived from such a setting. Instead, the sponsor reports the observed value of the nominal p-value process $(p_r)$ to the regulator ($p_{r_1}$ or $p_{r_2}$), which in turn compares this value with the adopted curve ($b_p(r)$ or $b^c_p(r)$) and accepts the new treatment $A$ if and only if the reported value is below the curve.

Consider now the scenario of a common disease. The parameter values chosen implies that $b^c_p(r_1) > b_p(r_1)$, so that the classical rule is less conservative than $b_p$. Therefore, if the new treatment is accepted when $b_p$ is used, it will also be accepted in the classical regime. The situation is reversed in the scenario of a rare disease, because in this case $b^c_p(r_2) < b_p(r_2)$. This means that there are possible outcomes of the trial which will result in acceptance of $A$ when $b_p$ is used but not when $b^c_p$ is used.

\subsection{On the two extensions}

\subsubsection{The asymmetric problem}
Assuming that $q > 0$ leads to asymmetric optimal stopping boundaries. The standardised version of the problem in equation \eqref{eq:problem_asym_standardised} can still be solved using the backwards trapezoidal method described in Section \ref{sec:solving_numerically}. However, because of the asymmetry, it is necessary to solve two coupled integral equations when applying the backwards procedure producing the piecewise linear approximations to the boundaries. We omit the details, since only minor modifications of the algorithm used when $q = 0$ are required.

It is easy to imagine practical applications in which the appropriate value for $q$ would be quite large. For example, consider a trial lasting one year and including one thousand patients, aimed at comparing a new candidate treatment $A$ with a standard alternative $B$ which is administered to one million patients outside the trial. This results in $q = 1000$. It is interesting to note (see Figure \ref{fig:asymmetric_boundaries}) that the upper boundary for the acceptance of the new treatment $A$ rather rapidly approaches the optimal boundary corresponding to the limit $q \to \infty$. One is led to conjecture that the optimal solution in this limit could serve as an adequate approximation even if $q$ is only moderately large. 

In Section \ref{subseq:model_applications}, we argued that a fixed significance level test is too conservative when compared to the optimal boundary obtained for Anscombe's original, symmetric problem (with $q = 0$). Similarly, since $b^+_S(r)$ decreases with increasing $q$ for each fixed $r$, it follows that the optimal boundary for the symmetric problem would also be too conservative in many practical situations.

\subsubsection{The random time horizon}
When generalising the model to allow for a random $N$, it was assumed that 
\begin{enumerate}
  \item The horizon $N$ is independent of the underlying process, i.e., the noise part of the Brownian motion, given the drift $\mu$,
  \item the horizon $N$ is independent of $\mu$, and,
  \item the only information obtained about $N$ when observing the sum process continuously up to time point $t$ is that either $N = t$ or $N > t$.
\end{enumerate}
The first two assumptions are essential in order to arrive at the optimal stopping problem in equation \eqref{eq:random_N_stopping_problem}, in which the form of the function $f(t)$ completely determines the effect that the distribution of $N$ has on the optimal solution. Recall that $N$ is interpreted as the time point at which both of the treatments $A$ and $B$ become obsolete and are replaced by another treatment. Even though it seems quite reasonable to assume that the value of $N$ should therefore be determined by the properties of the treatment replacing $A$ and $B$, and not the observations made in a preceding trial, situations in which this does not hold are still conceivable. For example, if it turns out that a new treatment $A$ is greatly superior to a standard alternative $B$, it may be argued that this should make it harder to develop a superceding treatment which improves further upon $A$. The density for $N$ would change during the trial, shifting to the right as the estimate for $\mu$ grows. A generalisation of the random horizon model along these lines would be an interesting topic for further research. If assumption 1 above is kept while 2 is dropped, $f(t)$ in Section \ref{sec:random_nb} should be replaced by $f_\mu(t)$, and $h_\nu$ in equation \eqref{eq:random_N_stopping_problem} would have to be redefined so as to also include the dependence of $N$ on $\mu$.

Keeping the independence assumptions 1 and 2, it would also be interesting to consider problems in which the distribution for $N$ evolves in a more complicated way. For instance, instead of assuming that a fixed distribution specified at time 0 is updated by just conditioning on $N > t$, $N$ could be defined as the jump time (from 1 to 0) of a binary stochastic process adapted to some filtration that is independent of the trial observations. The choice of this process would probably have to be done quite carefully in order to make the problem tractable using available optimal stopping theory, but such a generalisation might be a way to model more realistic belief dynamics about $N$.

\subsection{Limitations and further research}
One of the most basic assumptions in Anscombe's model is that of normally distributed responses. It is vital to assume this in order to arrive at the familiar frawework of optimal stopping for a Brownian motion. However, because of the central limit theorem, the model can be used to obtain at least approximately optimal boundaries in a wide range of applications. Another assumption made is that the sample variance $\sigma^2$ is known. In practice, $\sigma^2$ would quite likely need to be estimated as the response data accrues. It would certainly be an interesting topic for further research to see just how much of the framework that eventually remains if a prior is placed on $\sigma^2$ and then updated sequentially.

Introducing $q$ as an additional parameter makes Anscombe's model more flexible, but also leads to the practical problem of estimating its value. Since the problem is similar to that of estimating $N$, one is led to the idea of introducing a stochastic model also for $q$. With sufficient independence assumptions, it might still be possible to proceed with the analysis, although the authors have made no efforts in this direction.

A possibly contentious issue is the choice of the utility function to optimise. Anscombe's formulation leads to an aggregated utility that values each patient equally. A unit of health is valued the same regardless of whether it benefits the first patient recruited to the trial or patient number $N$. In particular, no discounting of future health benefits is implemented, even though some form of discounting is very common in alternative models found in the financial and health economics literature. We stress that this utilitarian flavour of the model should not be viewed as something necessary. Rather, it seems like a natural starting point before introducing more complicating factors that may bring the model closer to practical applications. However, it is important to keep in mind that the definition of the utility function will have a large impact on which boundaries that are optimal.

Finally, we note that the numerical methods described in Section \ref{sec:solving_numerically} can certainly be improved upon. A more comprehensive investigation regarding which algorithms are best suited for different types of free-boundary value problems would be valuable. 

\newpage

\appendix

  \section{Proof of Theorem \ref{thm:integral_equation_uniqueness}} \label{sec:integral_equation_uniqueness}
\begin{proof}
  Uniqueness will be demonstrated by showing that if $\bar{b}_S(r) \ge 0$ solves the integral equation \eqref{eq:symmetric_free_boundary_equation_reduced}, then $\bar{b}_S(r) = b_S(r)$ for $r \in [0, 1]$. 

Define a value function corresponding to $\bar{b}_S$ as
\begin{equation} \label{eq:integral_eq_uniqueness_barV_def}
  \bar{V} (r, y) = \int_r^1 \! \mathbb{E}_{(r, y)} \left[ h_\xi (u, |S_u|) \, \mathbb{I} \left( | S_u | \ge \bar{b}_S (u) \right) \right] \, \mathrm{d} u , \quad 0 \le r \le 1.
\end{equation}
By comparing the right hand sides of equations \eqref{eq:symmetric_free_boundary_equation_reduced} and \eqref{eq:integral_eq_uniqueness_barV_def}, it is then clear that if $\bar{b}_S$ solves equation \eqref{eq:symmetric_free_boundary_equation_reduced},
\begin{equation}  \label{eq:integral_eq_uniqueness_barV_G}
\bar{V} (r, \pm \bar{b}_S(r)) = (1 - r) h_\xi (r, \bar{b}_S(r)) = \tilde{G}(r, \bar{b}_S(r)), \quad r \in [0, 1].
\end{equation}
Moreover, since the process $(r, S_r)_r$ has the strong Markov property, it follows that for any stopping time $\rho \in [r, 1]$,
\begin{equation} \label{eq:integral_eq_uniqueness_barV_stopping_time}
\bar{V} (r, y) = \mathbb{E}_{(r, y)} \left[ \bar{V} (\rho, S_\rho) \right] + \mathbb{E}_{(r, y)} \left[ \int_r^\rho \!  h_\xi (u, |S_u|) \, \mathbb{I} \left( | S_u | \ge \bar{b}_S (u) \right) \, \mathrm{d} u \right].
\end{equation}
That $\bar{b}_S(r) = b_S(r)$, $0 \le r \le 1$, will now be shown in four steps.\\

\noindent \underline{Step 1 ($\bar{V} (r, y) = \tilde{G} (r, y)$ if $|y| \ge \bar{b}_S(r)$):} \\
Let $|y| \ge \bar{b}_S(r)$ and $\rho = \inf \{u \ge r : |S_u| \le \bar{b}_S(u) \} \wedge 1$. This definition of $\rho$ ensures that $\rho \in [r, 1]$ and that the process $(S_r)$, when started in $(r, y)$, will always stay above $\bar{b}_S(u)$ for $u \in [r, \rho]$. Equations \eqref{eq:integral_eq_uniqueness_barV_G} and \eqref{eq:integral_eq_uniqueness_barV_stopping_time} now yield
\begin{align*}
  \bar{V} (r, y) & = \mathbb{E}_{(r, y)} \left[ \bar{V} (\rho, S_\rho) \right] + \mathbb{E}_{(r, y)} \left[ \int_r^\rho \! h_\xi (u, |S_u|) \, \mathbb{I} \left( | S_u | \ge \bar{b}_S (u) \right) \, \mathrm{d} u \right] \\
  & = \mathbb{E}_{(r, y)} \left[ \tilde{G} (\rho, S_\rho) \right] + \mathbb{E}_{(r, y)} \left[ \int_r^\rho \! h_\xi (u, |S_u|) \, \mathrm{d} u \right] \\
  & = \tilde{G} (r, y),
\end{align*}
where the last equality follows by Dynkin's formula applied to $\tilde{G} (\rho, S_\rho)$.\\

\noindent \underline{Step 2 ($\bar{V} (r, y) \le \tilde{V} (r, y)$):} \\
Since $\bar{V} (r, y) = \tilde{G} (r, y) \le \tilde{V} (r, y)$ if $|y| \ge \bar{b}_S(r)$ by step 1, it remains to show the inequality under the assumption that $|y| < \bar{b}_S(r)$. Applying equation \eqref{eq:integral_eq_uniqueness_barV_stopping_time} to the stopping time $\rho = \inf \{u \ge r : |S_u| \ge \bar{b}_S(u) \} \wedge 1$ gives
\begin{align*}
  \bar{V} (r, y) & = \mathbb{E}_{(r, y)} \left[ \bar{V} (\rho, S_\rho) \right] + \mathbb{E}_{(r, y)} \left[ \int_r^\rho \! h_\xi (u, |S_u|) \, \mathbb{I} \left( | S_u | \ge \bar{b}_S (u) \right) \, \mathrm{d} u \right] \\
  & = \mathbb{E}_{(r, y)} \left[ \tilde{G} (\rho, S_\rho) \right] + \mathbb{E}_{(r, y)} \left[ \int_r^\rho \! h_\xi (u, |S_u|) \, 0 \, \mathrm{d} u \right] \\
  & \le \tilde{V} (r, y) .
\end{align*}

\newpage

\noindent \underline{Step 3 $(\bar{b}_S \le b_S)$:} \\
Let $y \ge \max(b_S(r), \bar{b}_S(r))$. Using steps 1 and 2 and applying equation \eqref{eq:integral_eq_uniqueness_barV_stopping_time} to the stopping time $\rho = \inf \{u \ge r : |S_u| \le b_S(u) \} \wedge 1$ yields
\begin{align*}
  \tilde{G} (r, y) = \bar{V} (r, y) & = \mathbb{E}_{(r, y)} \left[ \bar{V} (\rho, S_\rho) \right] + \mathbb{E}_{(r, y)} \left[ \int_r^\rho \! h_\xi (u, |S_u|) \, \mathbb{I} \left( | S_u | \ge \bar{b}_S (u) \right) \, \mathrm{d} u \right] \\
  & \le \mathbb{E}_{(r, y)} \left[ \tilde{V} (\rho, S_\rho) \right] + \mathbb{E}_{(r, y)} \left[ \int_r^\rho \! h_\xi (u, |S_u|) \, \mathbb{I} \left( | S_u | \ge \bar{b}_S (u) \right) \, \mathrm{d} u \right].
\end{align*}
Similarly, the value function $\tilde{V}$ satisfies
\begin{align*}
  \tilde{G} (r, y) = \tilde{V} (r, y) & = \mathbb{E}_{(r, y)} \left[ \tilde{V} (\rho, S_\rho) \right] + \mathbb{E}_{(r, y)} \left[ \int_r^\rho \! h_\xi (u, |S_u|) \, \mathbb{I} \left( | S_u | \ge b_S (u) \right) \, \mathrm{d} u \right] \\
  & = \mathbb{E}_{(r, y)} \left[ \tilde{V} (\rho, S_\rho) \right] + \mathbb{E}_{(r, y)} \left[ \int_r^\rho \! h_\xi (u, |S_u|) \, \mathrm{d} u \right].
\end{align*}
It follows that
\begin{equation*}
  \mathbb{E}_{(r, y)} \left[ \int_r^\rho \! h_\xi (u, |S_u|) \left( 1 - \mathbb{I} \left( | S_u | \ge \bar{b}_S (u) \right) \right) \, \mathrm{d} u \right] \le 0 ,
\end{equation*}
and hence
\begin{equation*}
  \mathbb{P}_{(r, y)} \left( | S_u | \ge \bar{b}_S (u) \text{ for all $u$ such that } r \le u \le \rho \right) = 1 .
\end{equation*}
Since both $\bar{b}_S$ and $b_S$ are assumed to be continuous, this can only be the case if $\bar{b}_S \le b_S$. \\

\noindent \underline{Step 4 $(\bar{b}_S \ge b_S)$:} \\
Using step 2 and applying equation \eqref{eq:integral_eq_uniqueness_barV_stopping_time} to the stopping time $\rho = \inf \{u \ge r : |S_u| \ge b_S(u) \} \wedge 1$ (which is the optimal one for the original problem) yields
\begin{equation*}
\tilde{V} (r, y) \ge \bar{V} (r, y) = \mathbb{E}_{(r, y)} \left[ \bar{V} (\rho, S_\rho) \right] + \mathbb{E}_{(r, y)} \left[ \int_r^\rho \! h_\xi (u, |S_u|) \, \mathbb{I} \left( | S_u | \ge \bar{b}_S (u) \right) \, \mathrm{d} u \right] .
\end{equation*}
Since $|S_\rho| = b_S( \rho ) \ge \bar{b}_S ( \rho )$ by step 3, step 1 implies that $\bar{V} (\rho, S_\rho) = \tilde{G} (\rho, S_\rho)$, so that
\begin{equation*}
  \tilde{V} (r, y) \ge \tilde{V} (r, y) + \mathbb{E}_{(r, y)} \left[ \int_r^\rho \! h_\xi (u, |S_u|) \, \mathbb{I} \left( | S_u | \ge \bar{b}_S (u) \right) \, \mathrm{d} u  \right].
\end{equation*}
The above implies that
\begin{equation*}
  \mathbb{P}_{(r, y)} \left( | S_u | \le \bar{b}_S (u) \text{ for all $u$ such that } r \le u \le \rho \right) = 1 .
\end{equation*}
Since both $\bar{b}_S$ and $b_S$ are assumed to be continuous, this can only hold if $\bar{b}_S \ge b_S$.
\end{proof}
\newpage
\section{Details on the caluclations for Theorem \ref{thm:asymptotic_expansion}} \label{sec:asymptotic_expansion}
    Let $\hat{V}_0 (s, y)$ be the value function corresponding to the modified reward function $\hat{G}_0$ and define $z = z(s) = c_q^+ (s) / \sqrt{-s}$. The boundary conditions of continuous and smooth fit which $\hat{V}_0$ must satisfy on $c_q^+ (s)$ are
  \begin{align*}
    \hat{V}_0 & = \hat{G}_0 = -z (-s)^{-1/2} (1 + 2q) - 2 (-s)^{1/2} (1 + q) \big( \phi (z) - z \left(1 - \Phi(z) \right) \big), \\
    \partial_{y} \hat{V}_0 & = \partial_y \hat{G}_0 = -(-s)^{-1} (1 + 2q) + 2 (1 + q) \left( 1 - \Phi(z) \right).
  \end{align*}
  Setting $\eta = (1 + q) / (1 + 2q)$, these may be rewritten as 
  \begin{align}
    \frac{\hat{V}_0 (s, c_q^+ (s))}{1 + 2q} & = -z (-s)^{-1/2} - 2 (-s)^{1/2} \eta \big( \phi (z) - z \left(1 - \Phi(z) \right) \big), \label{eq:asymmetric_expansion_BC1} \\
    \frac{\partial_{y} \hat{V}_0 (s, c_q^+ (s))}{1 + 2q} & = -(-s)^{-1} + 2 \eta \left( 1 - \Phi(z) \right). \label{eq:asymmetric_expansion_BC2}
  \end{align}  
  Following \citet[Appendix A5]{Chernoff1981}, we assume that $c_q^+ (s)$ satisfies 
  \begin{equation}
    \ln (-s) = z^2 / 2 + a_{-1} \ln (z) + a_0 + o(1), \quad s \to -\infty, \label{eq:boundary_expansion}
  \end{equation}
  for some constants $a_{-1}$ and $a_0$ (to be determined), and that
  \begin{equation*}
    \frac{\hat{V}_0 (s, c_q^+ (s))}{1 + 2q} \sim f_0 \left( (-s)^{1/2} z \right), \quad \frac{\partial_{y} \hat{V}_0 (s, c_q^+ (s))}{1 + 2q} \sim f_0' \left( (-s)^{1/2} z \right), \quad s \to -\infty .
  \end{equation*}
  The function $f_0$ is defined as
  \begin{equation*}
    f_0 (y) = - \frac{2  \ln (y^2)}{y}, \quad \text{giving  } f_0' (y) = 2 \left( \frac{\ln (y^2) - 2}{y^2} \right).
  \end{equation*}
  In what follows, we'll make use the well known asymptotic results for the normal density and distribution functions given below:
  \begin{align}
    1 - \Phi(z) & = \phi(z) \left( z^{-1} + o(z^{-1}) \right), \quad z \to \infty, \label{eq:normal_cdf_exp1} \\ 
    \phi (z) - z(1 - \Phi(z)) & = \phi(z) \left( z^{-2} + o(z^{-2}) \right), \quad z \to \infty. \label{eq:normal_cdf_exp2}
  \end{align}
  Use of equation \eqref{eq:normal_cdf_exp2} in equation \eqref{eq:asymmetric_expansion_BC1} now gives
  \begin{align*}
    -2 \frac{\ln (-s) + \ln (z^2)}{(-s)^{1/2} z} & \sim -z (-s)^{-1/2} - 2 (-s)^{1/2} \eta \phi(z) \left( z^{-2} + o(z^{-2}) \right) \iff \\
    \frac{-2 z^{-1}}{(-s)^{1/2}} \left( \ln (-s) + \ln (z^2) \right) & \sim \frac{- 2 z^{-1}}{(-s)^{1/2}} \left( z^2 / 2 + (-s) \eta \phi(z) \left( z^{-1} + o(z^{-1}) \right) \right) .
  \end{align*}
  $\ln (-s) + \ln (z^2)$ should now be matched against $z^2 / 2 + (-s) \eta \phi(z) \left( z^{-1} + o(z^{-1}) \right)$. Insertion of equation \eqref{eq:boundary_expansion} into these expressions leads to
  \begin{align}
    \ln (-s) + \ln (z^2) & = z^2 / 2 + a_{-1} \ln (z) + a_0 + o(1) + \ln (z^2), \label{eq:asymmetric_match11} \\
    z^2 / 2 + (-s) \eta \phi(z) \left( z^{-1} + o(z^{-1}) \right) & = z^2 / 2 + \frac{z^{a_{-1} - 1} e^{a_0} e^{o(1)} \eta \left( 1 + o(1) \right)}{\sqrt{2\pi}} \label{eq:asymmetric_match12}.
  \end{align}  
  Similarly, use of equation \eqref{eq:normal_cdf_exp1} in equation \eqref{eq:asymmetric_expansion_BC2} gives
  \begin{align*}
    2 \left( \frac{\ln (-s) + \ln (z^2) - 2}{(-s) z^2 } \right) & \sim -(-s)^{-1} + 2 \eta \phi(z) \left( z^{-1} + o(z^{-1}) \right) \iff \\
    \frac{2 z^{-2}}{(-s)} \left( \ln (-s) + \ln (z^2) - 2 \right) & \sim \frac{2 z^{-2}}{(-s)} \left( (-z^2 / 2) + (-s) z^2 \eta \phi(z) \left( z^{-1} + o(z^{-1}) \right) \right),
  \end{align*}
  and $\ln (-s) + \ln (z^2) - 2$ should be matched against $(-z^2 / 2) + (-s) z^2 \eta \phi(z) \left( z^{-1} + o(z^{-1}) \right)$. Insertion of equation \eqref{eq:boundary_expansion} on both sides leads to
  \begin{align}
    \ln (-s) + \ln (z^2) - 2 & = z^2 / 2 + a_{-1} \ln (z) + a_0 + o(1) + \ln (z^2) - 2, \label{eq:asymmetric_match21} \\
    (-z^2 / 2) + (-s) z^2 \eta \phi(z) \left( z^{-1} + o(z^{-1}) \right) & = (-z^2/2) + \eta \frac{z^{a_{-1} + 1} e^{a_0} e^{o(1)} \left( 1 + o(1) \right)}{\sqrt{2 \pi}} \label{eq:asymmetric_match22} .
  \end{align}
  By choosing $a_{-1} = 1$ and $a_0 = (1/2) \ln (2 \pi) + \ln (\eta^{-1})$, the coefficients of the dominating terms (involving $z^2$) on the right hand sides of equations \eqref{eq:asymmetric_match11} and \eqref{eq:asymmetric_match12}, and that of equations \eqref{eq:asymmetric_match21} and \eqref{eq:asymmetric_match22}, will agree. Insertion of these coefficient values in equation \eqref{eq:boundary_expansion} leads to
  \begin{equation*}
    \frac{ \phi \left( c_q^+ / \sqrt{-s} \right) }{ c_q^+ / \sqrt{-s} } \sim \eta^{-1} (-s)^{-1}, \quad s \to -\infty .
  \end{equation*}
  By the assumption \eqref{eq:boundary_expansion}, as $s \to -\infty$, $c_q^+ / \sqrt{-s} \to \infty$. Hence, an application of equation \eqref{eq:normal_cdf_exp1} yields
  \begin{equation*}
    1 - \Phi(c_q^+ / \sqrt{-s} ) \sim \eta^{-1} (-s)^{-1}, \quad s \to -\infty ,
  \end{equation*}
  which in turn implies
  \begin{equation*}
    c_q^+ = \sqrt{-s} \, \Phi^{-1} \left( 1 + \eta^{-1} s^{-1}(1 + o(1)) \right), \quad s \to -\infty.
  \end{equation*}  
  From this point it is easily shown, using the following asymptotic properties for the normal inverse distribution function and the Lambert W function \citep{Dominici2003,Corless1996},
  \begin{equation*}
    \Phi^{-1} (1 - x) \sim \sqrt{W \left( \frac{1}{2 \pi x^2} \right)}, \quad x \to 0, \qquad  W(x) \sim \ln \left( \frac{x}{\ln (x)} \right), \quad x \to \infty,
  \end{equation*}
  that we may in fact write $c_q^+ \sim \sqrt{-s} \, \Phi^{-1} \left( 1 + \eta^{-1} s^{-1} \right), \,  s \to -\infty$.

\newpage

\small \selectfont

\bibliography{freq_math}

\end{document}